\documentclass[reprint,aps,prb,twocolumn,superscriptaddress,showpacs]{revtex4-1}

\usepackage{graphicx}
\usepackage{dcolumn}
\usepackage{bm}
\usepackage{bbold}
\usepackage{natbib}
\usepackage{amsmath}
\usepackage{color}

\begin{document}

\title{Aharonov-Bohm and Aharonov-Casher effects for local and nonlocal Cooper pairs}

\author{Damian Tomaszewski}
\affiliation{Institute of Molecular Physics, Polish Academy of Science, Smoluchowskiego 17, 60-179 Poznan, Poland}
\author{Piotr Busz}
\affiliation{Institute of Molecular Physics, Polish Academy of Science, Smoluchowskiego 17, 60-179 Poznan, Poland}
\author{Rosa L\'opez}
\affiliation{Institut de F\'{\i}sica Interdisciplin\`aria i Sistemes Complexos IFISC (CSIC-UIB), E-07122 Palma de Mallorca, Spain}
\author{Rok \v{Z}itko}
\affiliation{Jo\v{z}ef Stefan Institute, Jamova 39, SI-1000 Ljubljana, Slovenia}
\affiliation{Faculty  of Mathematics and Physics, University of Ljubljana, Jadranska 19, SI-1000 Ljubljana, Slovenia}
\author{Minchul Lee}
\affiliation{Department of Applied Physics, College of Applied Science, Kyung Hee University, Yongin 446-701, Korea}
\author{Jan Martinek}
\affiliation{Institute of Molecular Physics, Polish Academy of Science, Smoluchowskiego 17, 60-179 Poznan, Poland}

\date{\today}

\begin{abstract}
We study combined interference effects
due to the Aharonov-Bohm (AB) and Aharonov-Casher (AC) phases
in a Josephson supercurrent of local and nonlocal (split) Cooper pairs.
We analyze a junction between two superconductors interconnected through a normal-state nanostructure with either
(i) a ring, where single-electron interference is possible,
or (ii) two parallel nanowires, where the single-electron interference can be absent,
but the cross Andreeev reflection can occur.
In the low-transmission regime in both geometries
the AB and AC effects can be related to only local or nonlocal Cooper pair transport, respectively.
\end{abstract}

\flushbottom
\maketitle

\thispagestyle{empty}

\section{Introduction}

Substantial progress has been made in recent years
in the creation of spatially separated spin-entangled electrons in solid state
by Cooper pair splitting~\cite{RecherPRB01,EldridgePRB10,HofstetterNature2009,HerrmannPRL10,SchindelePRL12,DasNC2012,TanPRL2015}.
Such entangled states are a necessary ingredient of quantum communication and computing~\cite{NielsenQI2000}.
It has been also demonstrated that a Josephson supercurrent with unusual properties
can be generated from nonlocal split Cooper pairs~\cite{DeaconNC15},
as pointed out by Wang and Hu~\cite{WangPRL11} in regard to the Aharonov-Bohm (AB) effect.
This new Josephson current requires further studies, in particular of its interference properties.

One of the best-known interference phenomena  is the Aharonov-Bohm (AB) effect~\cite{AharonovPR59,SharvinJETP81,WebbPRL85,vanOudenaardenNature98},
where the phase of a charged particle is affected by magnetic flux.
Dual to the AB phenomenon is the Aharonov-Casher (AC) effect~\cite{AharonovPRL84,CimminoPRL89,Mathur.68.2964},
in which electric field acts on the phase of magnetic moment.

The AC effect for electrons in solid state can be caused for instance by the Rashba spin-orbit interaction,
observed in mesoscopic rings~\cite{KonigPRL06,BergstenPRL06},
or in the Datta-Das transistor~\cite{DattaAPL90,KooScience2009},
where oscillations of conductance as a function of electric field occur
due to the Rashba phase~$\phi_R$.
Such interaction is of major importance for spintronics,
because its strength can be controlled by an external  gate voltage.

In s-wave superconductors, the Cooper pairs are in the singlet state, and thus have no net magnetic moment (spin $S = 0$). Therefore, it was recently postulated that there should be no AC effect for such a composite object.
This conjecture can be also linked to the fact that the two spin components ($\sigma=\pm1$ for spin $\uparrow, \downarrow$) of a Cooper pair in a quasi-1D quantum wire
have opposite Rashba phases $\sigma \phi_R$~\cite{RashbaSPSS60,BychkovJPC84,ManchonNatMat2015,RSOI}, which cancel each other and suppress the AC effect. Accordingly, it has been shown in a number of papers that to achieve modification of the Josephson current by the spin-orbit interaction one needs breaking of the time-reversal symmetry,
e.g. by a magnetic-field-induced Zeeman splitting or by magnetic exchange interactions \cite{BezuglyiPRB02,KriveLTP04,KrivePRB05,DellAnnaPRB07,BuzdinPRL08,ReynosoPRL08,ZazunovPRL09,BrunettiPRB13,YokoyamaPRB14,SamokhvalovSR14,JacobsenPRB15,MironovPRL15,CampagnanoJPCM15}.
We show that the desired spin control without any magnetic field can be achieved for split nonlocal Cooper pairs.

As a Cooper pair is composed of two electrons -- each of them having a magnetic moment related to its spin ($S = 1/2$) -- one may raise a question whether it is possible to induce the AC effect for each electron of a pair separately so that the two contributions do not compensate each other. Our answer to this question is positive, but only if  a Cooper pair is split and nonlocally preserves its entangled singlet state, while each electron of the pair experiences a different Rashba phase. The effect does not depend on the detailed geometry of the device as we prove by considering different cases. In all we find that at low transmission, $T \ll 1$,
the AB and AC effects are linked to
local~\cite{WangPRL11} and split nonlocal Cooper pair transport, respectively.
This explains why the AC effect has not been found for local Cooper pairs without breaking the time-reversal symmetry
in Refs.~\cite{BezuglyiPRB02,KriveLTP04,KrivePRB05,DellAnnaPRB07,ZazunovPRL09,BrunettiPRB13,YokoyamaPRB14,
ReynosoPRL08,BuzdinPRL08,MironovPRL15,CampagnanoJPCM15,JacobsenPRB15,SamokhvalovSR14},
and opens the possibility to control the two components
of the Josephson current
independently by the respective phases.

\begin{figure}[ht!]
\centering
\includegraphics{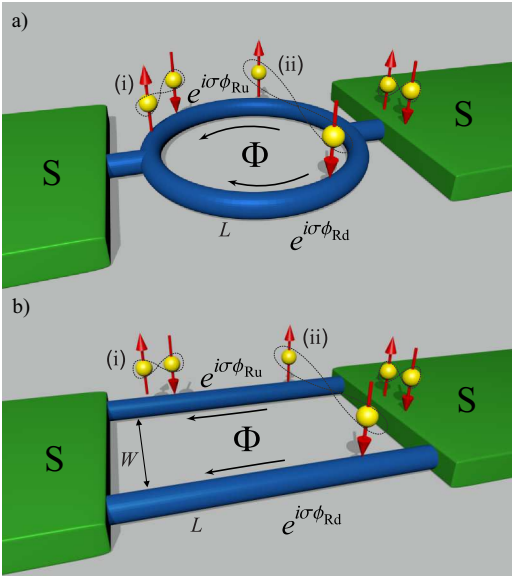}
\caption{Superconducting leads connected by a semiconducting link in the form of (a) a 1D ring, (b) two nanowires. Electrons flowing in the system acquire phases $\phi_{\rm{AB}}$ and $\sigma\phi_{\rm{Ru/d}}$ related to the external magnetic field flux~$\Phi$ and the Rashba spin-orbit interaction, respectively. In both systems local (i) and nonlocal (ii) Cooper pair transport is possible.}
\label{fig:scheme}
\end{figure}

Below we consider two different setups, with two superconducting electrodes linked by:
(i) a normal 1D ring, in which single-electron interference is possible (FIG.~\ref{fig:scheme}(a));
(ii) two parallel nanowires (2NW) (FIG.~\ref{fig:scheme}(b)),
where single-electron interference can be absent, but cross Andreeev reflection (CAR) is possible (the distance between the nanowires is comparable to or smaller than the Cooper pair size~$\xi$).

\section{Junction with ring}

In the first case to be considered the superconducting leads are connected by a 1D ring
formed by two Y-junctions and two arms (up and down).
We assume that the size~$L$ of the system
is smaller than the phase coherence length~$l_\phi$, $L<l_\phi$,
which implies the possibility of single electron quantum interference in a normal state.

The Josephson current for $k_B T\rightarrow 0$ and $L\ll\xi_0$ can be calculated from the equation~\cite{BeenakkerPRL91,BagwellPRB92}:
\begin{equation}
{I}\left( \varphi  \right) = \frac{{2\rm{e}}}{\hbar }\frac{\partial }{{\partial \varphi }}\sum\limits_n {{E_{n-} }\left( \varphi  \right)}\;,
\label{eqn:JosephsonCurrent}
\end{equation}
where the sum runs over all negative Andreev bound states energies,
which can be calculated from Beenakker's determinant equation, using scattering matrix formalism~\cite{BeenakkerPRL91,LeeNatureNano2013,WoerkomArxiv2016}:
\begin{align}
\textrm{Det} \left( {I - {\alpha ^2}r^*_{\rm{A}}{S_{\rm{e}}}r_{\rm{A}}{S_{\rm{h}}}} \right) = 0\;,
&\;{r_A} = \left( {\begin{array}{*{20}{c}}
e^{i\frac{\varphi}{2}}&0\\
0&e^{-i\frac{\varphi}{2}}
\end{array}} \right)\;,
\label{eqn:AndreevBoundStates}
\end{align}
where $\alpha  = \exp \left( { - i\arccos \left( {E/\Delta} \right)} \right)$, $r_A$ is the Andreev reflection matrix, with $\varphi$ denoting the superconducting phase difference and $S_{\rm{e/h}}$ is scattering matrix for electrons/holes.

The ring can be characterized by a scattering matrix (S-matrix) $S_{\rm{e}}$ (see APPENDIX A for details), with the parameter~$t_1$ describing the symmetric transmission
between the incoming electrode and each arm of the ring: \mbox{$0\le t_1\le1/\sqrt{2}$}, and where phase dependent transmission amplitudes of the up (down) arm are given by:
\begin{align}
{t_{\rm{u}\sigma /\rm{d}\sigma }} &= \exp \left( {i\left( {{\chi _{\rm{u/d}}} - \sigma {\phi _{\rm{Ru/d}}} \mp \frac{{{\phi _{\rm{AB}}}}}{2}} \right)} \right)\;,
\label{eqn:tud}\\
{t'_{ \rm{u} \sigma/\rm{d} \sigma}}& = \exp \left( {i\left( {{\chi _{\rm{u/d}}} + \sigma {\phi _{\rm{Ru/d}}} \pm \frac{{{\phi _{\rm{AB}}}}}{2}} \right)} \right)\;, \nonumber
\end{align}
where the subscript~$\rm{u}$ ($\rm{d}$) indicates the up (down) arm and the prime denotes the transmission in opposite direction.
Here $\chi_{\rm{u/d}}$ denote the respective dynamic phases~\cite{NazarovQT}, that have the same sign for all cases, while
$\sigma\phi_{\rm{Ru/d}}$ are the spin-dependent Rashba phases,
and $\phi_{\rm{AB}}=\pi \Phi/\Phi_0$ is the AB phase, with $\Phi_0=\pi \hbar \rm{c/e}$, which both switch signs while changing direction and the AB phase has opposite sign for two arms~\cite{ABAsy}.
In further calculations we assume for simplicity ~$\chi_{\rm{u}}=\chi_{\rm{d}}=\pi/2$; this does not affect the qualitative validity of the conclusions.
We consider the short SNS junction limit, $L \ll \xi_0=\hbar v_F/\Delta\left(0\right)$,
in which the scattering matrix $S_{\rm{e}}$ is independent of energy~\cite{BeenakkerPRL91}.
The hole S\nobreakdash-matrix~$S_{\rm{h}}$ is related to the electron S\nobreakdash-matrix~$S_{\rm{e}}$,
which is now spin-dependent, as $S_{\rm{h}}=\mathcal{T} S_{\rm{e}} \mathcal{T}^{-1}$,
where $\mathcal{T}=i \sigma_y K$, $\sigma_y$ denotes the Pauli matrix acting on the spin degree of freedom,
and $K$ is the operator of complex conjugation.
This implies $S_{\rm{h}\sigma}=S^*_{\rm{e}\bar \sigma}$~\cite{UnitCond}.

By solving Eq.~(\ref{eqn:AndreevBoundStates})
we obtain the bound state energy,
which is spin-independent for the particle-hole symmetry:
\begin{align}
E_\pm &=\pm \Delta \sqrt {\frac{{1 + \sqrt{{R_\uparrow }{R_{\downarrow }}} + {\mathop{\rm sgn}} \left(\Omega_1\right) \sqrt {{T_\uparrow T_{\downarrow }}} \cos \varphi }}{2}} \;,
\label{eqn:energyRing}\\
\Omega_1&=\frac{1}{2}\left[\cos {{\phi _{\rm{AB}}}} + \cos \left( {{\phi _{\rm{Ru}}} - {\phi _{\rm{Rd}}}} \right)\right]\;,\nonumber
\end{align}
where $R_\sigma=1-T_\sigma$, and $T_\sigma$ is the spin-dependent transmission of the ring:
\begin{align}
T_\sigma &=\frac{{8{t_1^4}\left( {1 + \Theta_\sigma} \right)}}{{{{\left( {3 - 3{t_1^2} + {\tilde t_1}  + \left( {1 - {t_1^2} - {\tilde t_1} } \right)\Theta_\sigma} \right)}^2}}}\;,
\label{eqn:Tsigma}\\
\Theta_\sigma &=\cos \left(\phi _{\rm{AB}} + \sigma \left( \phi _{\rm{Ru}} - \phi _{\rm{Rd}}\right)\right)\;,\nonumber
\end{align}
with ${\tilde t_1}=(1-2t^2 _1)^{1/2}$. The transmission $T_\sigma$ in Eq.~(\ref{eqn:Tsigma}) depends on the AB and AC phases
through the term $\Theta_\sigma$, which is spin-dependent only when
both $\phi_{\rm{AB}} \neq 0$ and $\phi_{\rm{Ru}}-\phi_{\rm{Rd}} \neq 0$.
Equation~(\ref{eqn:energyRing}) implies that the Andreev bound state energy
cannot be expressed only in terms of normal transmission $T_\uparrow$, $T_\downarrow$.

By substituting $\phi_{\rm{AB}}=0$~\cite{NoteSymT} or $\phi_{\rm{Ru}}-\phi_{\rm{Rd}}=0$
to Eq.~(\ref{eqn:energyRing}) and putting $T_\uparrow=T_\downarrow\equiv T_0$
we obtain the well-known result for the Andreev bound state energy~\cite{BeenakkerPRL91}:
\begin{equation}
E_\pm= \pm \Delta \sqrt {1 - T_{0}{{\sin }^2}\left(\varphi /2\right)}\;.
\label{eqn:energyDef}
\end{equation}


For a junction with ring, at low transmission $T_\sigma \ll 1$ the Josephson current has the form:
\begin{equation}
{I}\left( \varphi  \right) =\frac{{\rm{e}\Delta}}{2\hbar }{\mathop{\rm sgn}} \left(\Omega_1\right)\sqrt{T_\uparrow T_\downarrow} \sin \varphi=\frac{{\rm{e}\Delta}}{2\hbar }{t^4 _1} \Omega_1 \sin \varphi \;.
\label{eqn:JosephsonCurrentLowestorderRing}
\end{equation}
The current~(\ref{eqn:JosephsonCurrentLowestorderRing}) has two components,
one dependent on the $\phi_{\rm{AB}}$ phase
and the other on the Rashba phase $\phi_{\rm{Ru}}-\phi_{\rm{Rd}}$
(see Eq.~(\ref{eqn:energyRing}) for~$\Omega_1$).
In the low-transmission regime, $T_\sigma \ll 1$, this dependence can be related to
the way Cooper pairs flow through the system.
If both electrons of a Cooper pair (in an $\left|S\right\rangle$ state)
travel in the same arm of the ring,
their Rashba phases cancel due to their opposite spins,
and the Josephson current only depends on the AB phase.
If a Cooper pair is split and the constituent electrons travel in different arms of the ring,
the AB phases of the electrons cancel, being opposite in the two arms;
consequently, this component of the Josephson current only depends on the Rashba phase, thus we can observe the AC effect.
In the higher-transmission regime more complex trajectories are available,
which prevents the separation of the two components.

\section{Junction with two nanowires}

We now show that the discussed effects do not depend on the geometry of the system.
We consider two nanowires connecting two superconducting electrodes (FIG.~\ref{fig:scheme}(b))
spaced by a distance~$W$ smaller than the size~$\xi$ of the Cooper pair, $W\lesssim \xi$,
which can be larger than $l_\phi$ in \emph{dirty} superconductors.
In such a system the CAR effect is possible even though
there can be no single-electron interference in the normal state,
especially at higher temperatures close to $T_c$, since $l_{\phi}\propto T^{-1/2}$,
while $\xi\simeq\xi_0$ and only slightly varies with temperature~\cite{WashburnAP86,MourachkineJS2004}.
The CAR probability is a function of both $\xi$ and the Fermi
wavelength $\lambda_F$~\cite{RecherPRB01,FalciEuroLet2001,OhPRB2005},
nonetheless, the CAR in parallel nanowires coupled to a single
superconductor was observed experimentally at a distance $W$ between nanowires from 100~nm to 800~nm~\cite{BeckmannPRL2004,CaddenZimanskyNatPhys2009,BrauerPRB2010}.
The S\nobreakdash-matrix~$S_{\rm{e}\sigma}$ of this 2NW system (see APPENDIX B for details)
is a combination of the S-matrices~$S_{\rm{u}\sigma/\rm{d}\sigma }$ of each nanowire,
where $\tau_{\rm{u} \sigma /\rm{d} \sigma}\equiv t_2 t_{\rm{u} \sigma /\rm{d} \sigma}$
and $\tau'_{\rm{u} \sigma /\rm{d} \sigma}\equiv t_2 t'_{\rm{u} \sigma /\rm{d} \sigma}$
are the transmission amplitudes through a single (up ($\rm{u}$) or down ($\rm{d}$)) wire,
with the parameter~$t_2$ ranging from 0 to 1, $0\le t_2\le1$.
We assume that the wires are symmetric in the transmission parameter~$t_2$~\cite{Assym}.
In this system, when an electron (hole) enters the superconductor,
a hole (electron) can be reflected to any of the two available wires.
This two-nanowire Andreev reflection can be modeled as follows:
\begin{align}
{r_{\rm{A}}} = \left( {\begin{array}{*{20}{c}}
r_{\rm{a}}&0\\
0&r^*_{\rm{a}}
\end{array}} \right),\;
&{r_{\rm{a}}} = \left( {\begin{array}{*{20}{c}}
\sqrt {1 - {\gamma ^2}} e^{i\frac{\varphi}{2}}&{\gamma e^{i\frac{\varphi}{2}} }\\
{\gamma e^{i\frac{\varphi}{2}} }&-{\sqrt {1 - {\gamma ^2}} e^{i\frac{\varphi}{2}} }
\end{array}}\right),
\label{eqn:rAMatrix2ch}
\end{align}
where $\gamma\in\left\langle0,1\right\rangle$ describes the mixing amplitude between the two wires.
The solution of Eq.~(\ref{eqn:AndreevBoundStates}) yields four Andreev bound state energies:
\begin{equation}
E_{ n \pm}  =  \pm \Delta \sqrt {\frac{2 - T\left(  1 - \Omega_2 \cos \varphi   + n  {\sin \varphi }\sqrt {1 - \Omega_2 ^2}\right) }{2}} \;,
\label{eqn:energy2ch}
\end{equation}
where $T=\tau_{\rm{u} \sigma /\rm{d} \sigma}\tau^*_{\rm{u} \sigma /\rm{d} \sigma}=t^2_2$
is the spin- and phase-independent transmission of a single wire, $n=\pm 1$, and:
\begin{equation}
\Omega_2  = \left( {1 - {\gamma ^2}} \right)\cos {{\phi _{\rm{AB}}}} + {\gamma ^2}\cos \left( {{\phi _{\rm{Ru}}} - {\phi _{\rm{Rd}}}} \right)\;.
\label{eqn:omega2}
\end{equation}
In extreme cases $\Omega_2 = \cos\phi_{\rm{AB}}$ for $\gamma = 0$ and $\Omega_2 = \cos(\phi_{\rm{Ru}}-\phi_{\rm{Rd}})$ for $\gamma = 1$.

\begin{figure}[ht!]
\centering
\includegraphics{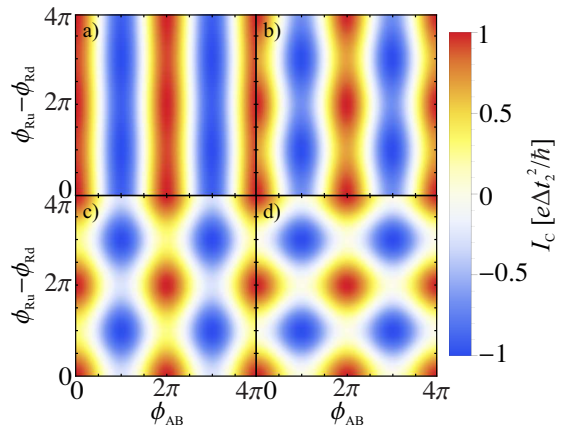}
\caption{Critical Josephson current in the 2NW system as a function of the AB and AC phases for different wire mixing: (a)~$\gamma=0.2$, (b)~$\gamma=0.4$, (c)~$\gamma=0.6$, (d)~$\gamma=1/\sqrt{2}$; $T \ll 1$.}
\label{fig:gamma}
\end{figure}

For low transmission, $T\ll 1$, the Josephson current is given by:
\begin{equation}
{I}\left( \varphi  \right) = \frac{{\rm{e}\Delta}}{\hbar } T \Omega_2 \sin \varphi \;.
\label{eqn:JosephsonCurrentLowestorder2ch}
\end{equation}
As in the ring system, also here the current has two components
related to different modes of electron pair flow (split or unsplit) through the system.
Comparing Eqs.~(\ref{eqn:JosephsonCurrentLowestorderRing}) and (\ref{eqn:JosephsonCurrentLowestorder2ch}),
we find that in the case of symmetric wire mixing, $\gamma=1/\sqrt{2}$,
the currents in the 2NW system and the ring system considered above
have the same phase dependence in the low-transmission regime, $T_\sigma\ll 1$.
When $\gamma\neq1/\sqrt{2}$, the 2NW system
has different amplitudes of AB and AC oscillations,
as indicated by Eq.~(\ref{eqn:omega2}) and illustrated by FIG.~\ref{fig:gamma}.
This is in contrast to the junction with ring, in which the amplitudes are equal.
In the extreme cases, for $\gamma=0$ (no mixing), Andreev bound states energies are given by:
\begin{equation}
E_{ n \pm}= \pm \Delta \sqrt{1-T \sin^2\frac{\varphi+n \phi_{\rm{AB}}}{2}} \;,
\label{eqn:energy2chGamma0}
\end{equation}
and for $\gamma=1$, in which case backscattering to the same wire is impossible and full splitting occurs:
\begin{equation}
E_{ n \pm}= \pm \Delta \sqrt{1-T \sin^2\frac{\varphi+n \sigma\left(\phi_{\rm{Ru}}-\phi_{\rm{Rd}}\right)}{2}} \;.
\label{eqn:energy2chGamma1}
\end{equation}
These specific situations can be regarded as
the flow of either unsplit or split Cooper pair electrons, respectively,
with the consequent dependence on only one phase (AB or AC).

\begin{figure}[ht!]
\centering
\includegraphics{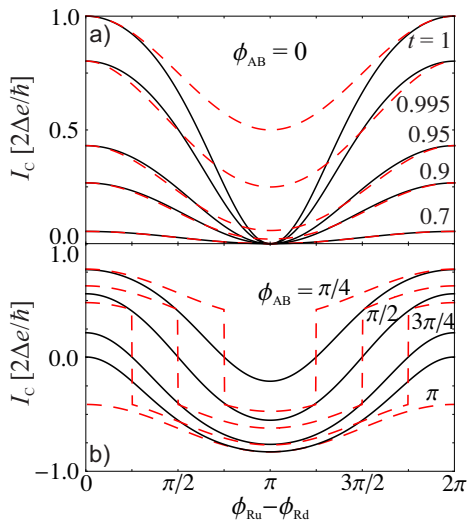}
\caption{Critical Josephson current $I_{\rm{C}}$ in ring (solid line) and 2NW (dashed line) junctions versus Rashba phase~$\phi_{\rm{Ru}}-\phi_{\rm{Rd}}$,
for (a) $t=t_1 \sqrt 2=\sqrt{2 t_2/\left(1+t_2\right)}$, $t\in\{0.7,0.9,0.95,0.995,1\}$, $\phi_{\rm{AB}}=0$; (b) $t=1$, $\phi_{\rm{AB}}=\{\frac{\pi}{4},\frac{\pi}{2},\frac{3\pi}{4},\pi\}$; see FIG.~\ref{fig:currents2D}(c,~d) for section along dashed line. Ring curves are multiplied by factor 2 since 2NW system consists effectively of two transport channels, while the ring has only one channel; $\gamma=1/\sqrt{2}$.}
\label{fig:currents}
\end{figure}

As we increase the junction transmission,
differences between these two systems become apparent also for $\gamma=1/\sqrt{2}$.
FIG.~\ref{fig:currents} shows the Josephson critical current~$I_{\rm{C}}$ plotted versus $\phi_{\rm{Ru}}-\phi_{\rm{Rd}}$
for $\phi_{\rm{AB}}=0$ and different values of parameter $t=t_1 \sqrt 2=\sqrt{2 t_2/\left(1+t_2\right)}$ ($0\le t\le 1$)~\cite{Paramt}.
In FIG.~\ref{fig:currents}(a) for $t \ll 1$
the characteristics are similar in the two systems.
A significant difference only occurs for large transmission, $t\approx 1$.
The same is observed in the current characteristics plotted for
different AB phases $\phi_{\rm{AB}}$ with $\phi_{\rm{Ru}}-\phi_{\rm{Rd}}=0$.

\begin{figure}[ht!]
\centering
\includegraphics{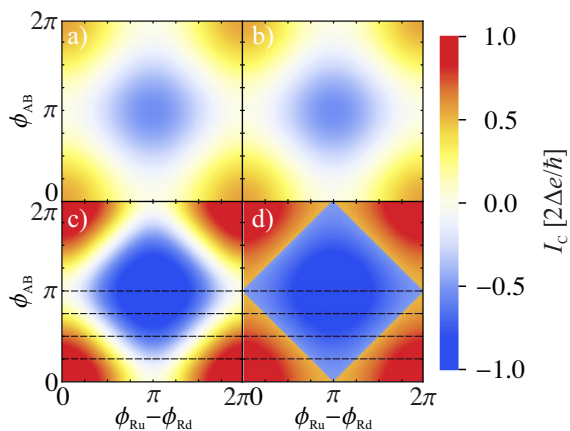}
\caption{Josephson critical current $I_{\rm{C}}$ versus AB phase~$\phi_{\rm{AB}}$ and Rashba phase~$\phi_{\rm{Ru}}-\phi_{\rm{Rd}}$ for \mbox{$t=0.7$}~(a,~b) and~1~(c,~d)
in junction with ring (a,~c) and in 2NW system (b,~d). Scaling: (a)~$\times 20$, (b)~$\times 10$, (c)~$\times 2$.}
\label{fig:currents2D}
\end{figure}

Another difference between the Josephson currents in these two systems
can be seen for $t\approx1$ when both phases are nonzero (FIG.~\ref{fig:currents}(b) and FIG.~\ref{fig:currents2D}).
In the 2NW system the current shows a step-like transition between positive and negative values
(see FIG.~\ref{fig:currents}(b)).
This is related to the fact that for the ring the transmission $T_\sigma$ depends on both AC and AB phases, therefore for a large range of parameters $T_\sigma < 1$. For the 2NW case the transmission of each nanowire does not depends on both phases. As a result the perfect transmission can be achieved, which make step-like behavior possible.

\section{Transmission amplitude asymmetry}

In the previous section, for simplicity, we consider symmetric two
nanowire system, however, experimental fabrication of junction with
two identically connected nanowires can be difficult. In this section
we prove that asymmetry in the transmission of two nanowires $t_2$, does not affect our main conclusions. In our
model we can introduce different amplitudes for up and down nanowire -
$t_{2\rm{u}}\neq t_{2\rm{d}}$ in S-matrix~Eq.~(\ref{eqn:Smatrix2ch}).
As a result the Josephson current for $T \ll 1$ has the form:
\begin{align}
{I}\left( \varphi  \right) =& I_{\rm{local}}\left(\varphi\right) + I_{\rm{nonlocal}}\left(\varphi\right)\;,\\
I_{\rm{local}} =&\frac{{\rm{e}\Delta}}{2\hbar}\left({1 - {\gamma ^2}} \right)\nonumber\\
&\times\left(t_{2\rm{u}}^2 \sin \left(\varphi + \phi _{\rm{AB}}\right)+t_{2\rm{d}}^2 \sin \left(\varphi - \phi _{\rm{AB}}\right)  \right)\;,\\
I_{\rm{nonlocal}} =&\frac{{\rm{e}\Delta}}{\hbar} t_{2\rm{u}}t_{2\rm{d}}{\gamma ^2}\cos \left( {{\phi _{\rm{Ru}}} - {\phi _{\rm{Rd}}}} \right)\sin \varphi \;.
\label{eqn:JosephsonCurrentLowestorder2chAssymetry}
\end{align}
The above equations confirm that transmission amplitude asymmetry does not change our general conclusion. The Josephson current, in low transmission regime, has two components as before: local - dependent only on the $\phi_{\rm{AB}}$ phase, which has two contributions from Cooper pairs flowing through up and down nanowire ($\propto t_{2\rm{u/d}}^2$), and the nonlocal component - dependent only on the Rashba phase $\phi_{\rm{Ru}} - \phi_{\rm{Rd}}$ ($\propto t_{2\rm{u}}t_{2\rm{d}}$).

\section{Conclusions}

We have demonstrated that the AC effect for Josephson supercurrent
is possible even in systems with unbroken time-reversal symmetry,
but only for nonlocal split Cooper pairs which can be free from the AB effect.
On the other hand, for local Cooper pairs the AC effect does not occur, while the AB effect has the standard form.
In the higher transmission regime, however, the local and nonlocal components will be mixed up by higher-order processes.
We have analyzed these effects in two different systems to show that discussed behavior is geometry independent.
One can expect a similar effects in Josephson junction with two parallel nanowires with a quantum dot inserted in each
nanowire~\cite{ArticleQD}.

In InAs and InSb nanowires a large spin-orbit coupling was observed with effective spin-orbit length $l_{\rm{so}}\approx200\rm{nm}$ and a Rashba parameter $\eta=0.2\rm{eV}\cdot\rm{{\AA}}$~\cite{FasthPRL2007,NadjPergePRL2012,MourikScience2012,ManchonNatMat2015}.
Recent experiment by S. Baba et al.~\cite{BabaArxiv2017} showed the
possibility of producing two Rashba parallel InAs nanowires
system with quantum dots (the length $\approx 250\rm{nm}$ and the
distance between nanowires $\approx 100\rm{nm}$). Experimental work by
D.B.~Szombati et al.~\cite{SzombatiNaturePhys2016} also shows
possibility of forming a Josephson junction with $\approx 200nm$
long InSb Rashba nanowire with quantum dot, with spin-orbit length
$l_{\rm{so}}\approx350\rm{nm}$, whereas S.~Gazibegovic et
al.~\cite{GazibegovicNature2017} show formation of InSb nanowire
"hashtags" (rectangular loops) that can be connected to
superconducting electrodes. The above examples of experimental work indicate that the proposed effects are possible to measure using present day technology.

\section*{Acknowledgements}

We would like to thank J.~Barna\'{s}, M.~Braun, B.~Braunecker, F.~Dominguez,
T.~Kontos, J.~K\"onig, D.~Loss, T.~Martin, C.~Sch\"{o}nenberger, B.~Sothmann, J.~Tworzyd{\l}o, and A.~L.~Yeyati for helpful discussions.
This study received support from the EU FP7 Project SE2ND (No. 271554) and the National Science Centre, Poland, grant 2015/17/B/ST3/02799.

\section*{Appendix A: S-matrix for ring}

In the first considered case the superconducting leads are connected by a 1D ring formed by two Y-junctions and two arms (up and down). Each part of the ring can be characterized by a scattering matrix (S-matrix). The left and right Y-junctions, with symmetric outputs, can be modeled by the following S\nobreakdash-matrices~\cite{Ouisse08}:
\begin{equation}
{S_{\rm{l}}} = \left( {\begin{array}{*{20}{c}}
{\tilde t_1}&t_1&t_1\\
t_1&{ - \frac{1}{2}\left( {1 + {\tilde t_1}} \right)}&{\frac{1}{2}\left( {1 - {\tilde t_1}} \right)}\\
t_1&{\frac{1}{2}\left( {1 - {\tilde t_1}} \right)}&{ - \frac{1}{2}\left( {1 + {\tilde t_1}} \right)}
\end{array}} \right)\;,
\label{eqn:smatrixL}
\end{equation}
\begin{equation}
{S_{\rm{r}}} = \left( {\begin{array}{*{20}{c}}
{ - \frac{1}{2}\left( {1 + {\tilde t_1}} \right)}&{\frac{1}{2}\left( {1 - {\tilde t_1}} \right)}&t_1\\
{\frac{1}{2}\left( {1 - {\tilde t_1}} \right)}&{ - \frac{1}{2}\left( {1 + {\tilde t_1}} \right)}&t_1\\
t_1&t_1&{\tilde t_1}
\end{array}} \right)\;,
\label{eqn:smatrixR}
\end{equation}
where ${\tilde t_1}=(1-2t^2 _1)^{1/2}$ and the parameter~$t_1$ describes the transmission
between the incoming electrode
and each arm of the ring: \mbox{$0\le t_1\le1/\sqrt{2}$}.
The central region of the ring, where an electron acquires a spin-dependent phase shift, can be described by two S-matrices:
\begin{align}
{S_{\rm{cu}\sigma/\rm{d}\sigma}}& = \left( {\begin{array}{*{20}{c}}
0&{{t'_{\rm{u}\sigma /\rm{d}\sigma }}}\\
{{t_{\rm{u}\sigma /\rm{d}\sigma }}}&0
\end{array}} \right)\;,
\label{eqn:smatrixC}\\
{t_{\rm{u}\sigma /\rm{d}\sigma }} &= \exp \left( {i\left( {{\chi _{\rm{u/d}}} - \sigma {\phi _{\rm{Ru/d}}} \mp \frac{{{\phi _{\rm{AB}}}}}{2}} \right)} \right)\;,
\label{eqn:tud}\\
{t'_{ \rm{u} \sigma/\rm{d} \sigma}}& = \exp \left( {i\left( {{\chi _{\rm{u/d}}} + \sigma {\phi _{\rm{Ru/d}}} \pm \frac{{{\phi _{\rm{AB}}}}}{2}} \right)} \right)\;,
\label{eqn:tudp}
\end{align}
where the subscript~$\rm{u}$ ($\rm{d}$) indicates the up (down) arm,
$\chi_{\rm{u/d}}$ are the respective dynamic phases~\cite{NazarovQT},
$\sigma\phi_{\rm{Ru/d}}$ the spin-dependent Rashba phases,
and $\phi_{\rm{AB}}=\pi \Phi/\Phi_0$ is the AB phase, with $\Phi_0=\pi \hbar \rm{c/e}$.
In our calculations we assume~$\chi_{\rm{u}}=\chi_{\rm{d}}=\pi/2$,
which corresponds to a particle-hole symmetry
that simplifies equations without loss of generality. The total scattering matrix~$S_{\rm{e}}$ for electrons passing through the ring
is a combination of matrices $S_{\rm{l}}$, $S_{\rm{r}}$ and $S_{\rm{cu}\sigma/\rm{d}\sigma}$:
\begin{equation}
{S_{\rm{e}\sigma,\rm{ring}}} = \left( {\begin{array}{*{20}{c}}
{{\rho _{\sigma,\rm{ring}}}}&{\tau {'_{\sigma,\rm{ring}}}}\\
{{\tau _{\sigma,\rm{ring}}}}&{{\rho _{\sigma,\rm{ring}}}}
\end{array}} \right)\;,
\label{eqn:SeRing}
\end{equation}
with:
\begin{widetext}
\begin{equation}
{\rho _{\sigma,\rm{ring}}} = \frac{{2{\tilde t_1} + \left( {1 - t_1^2 - {\tilde t_1}} \right)\left( {{t_{u\sigma }}t{'_{d\sigma }} + t{'_{u\sigma }}{t_{d\sigma }}} \right) - \left( {1 - t_1^2 + {\tilde t_1}} \right)\left( {{t_{u\sigma }}t{'_{u\sigma }} + {t_{d\sigma }}t{'_{d\sigma }}} \right) + 2{\tilde t_1}{t_{u\sigma }}t{'_{u\sigma }}{t_{d\sigma }}t{'_{d\sigma }}}}{{2 - \left( {1 - t_1^2 - {\tilde t_1}} \right)\left( {{t_{u\sigma }}t{'_{d\sigma }} + {t_{d\sigma }}t{'_{u\sigma }}} \right) - \left( {1 - t_1^2 + {\tilde t_1}} \right)\left( {{t_{u\sigma }}t{'_{u\sigma }} + {t_{d\sigma }}t{'_{d\sigma }}} \right) + 2{{\tilde t_1}^2}{t_{u\sigma }}t{'_{u\sigma }}{t_{d\sigma }}t{'_{d\sigma }}}}\;,
\label{eqn:rhoRing}
\end{equation}
\begin{equation}
{\tau _{\sigma,\rm{ring}}} = \frac{{4t_1^2\left( {{t_{u\sigma }} + {t_{d\sigma }} + \left( {t{'_{u\sigma }} + t{'_{d\sigma }}} \right){t_{u\sigma }}{t_{d\sigma }}} \right)}}{{4 - {{\left( {1 + {\tilde t_1}} \right)}^2}\left( {{t_{u\sigma }}t{'_{u\sigma }} + {t_{d\sigma }}t{'_{d\sigma }}} \right) - {{\left( {1 - {\tilde t_1}} \right)}^2}\left( {{t_{u\sigma }}t{'_{d\sigma }} + t{'_{u\sigma }}{t_{d\sigma }}} \right) + 4{{\tilde t_1}^2}{t_{u\sigma }}t{'_{d\sigma }}t{'_{u\sigma }}{t_{d\sigma }}}}\;,
\label{eqn:tauRing}
\end{equation}
\begin{equation}
\tau {'_{\sigma,\rm{ring}}} = \frac{{4t_1^2\left( {t{'_{u\sigma }} + t{'_{d\sigma }} - \left( {{t_{u\sigma }} + {t_{d\sigma }}} \right)t{'_{u\sigma }}t{'_{d\sigma }}} \right)}}{{4 - {{\left( {1 + {\tilde t_1}} \right)}^2}\left( {{t_{u\sigma }}t{'_{u\sigma }} + {t_{d\sigma }}t{'_{d\sigma }}} \right) - {{\left( {1 - {\tilde t_1}} \right)}^2}\left( {{t_{u\sigma }}t{'_{d\sigma }} + t{'_{u\sigma }}{t_{d\sigma }}} \right) + 4{{\tilde t_1}^2}{t_{u\sigma }}t{'_{d\sigma }}t{'_{u\sigma }}{t_{d\sigma }}}}\;.
\label{eqn:taupRing}
\end{equation}
\section*{Appendix B: S-matrix for two nanowires}
The S\nobreakdash-matrix~$S_{\rm{e}\sigma,\rm{2NW}}$ of the two parallel nanowires (2NW) system for $\chi_{\rm{u/d}}=\pi/2$ has the form:
\begin{equation}
{S_{\rm{e}\sigma,\rm{2NW}}} = \left( {\begin{array}{*{20}{c}}
\rho_{\rm{2NW}}&0&\tau'_{\rm{u} \sigma}&0\\
0&\rho_{\rm{2NW}}&0&\tau'_{\rm{d} \sigma}\\
\tau_{\rm{u} \sigma} &0& \rho_{\rm{2NW}}&0\\
0&\tau_{\rm{d} \sigma} &0& \rho_{\rm{2NW}}\\
\end{array}} \right),
\label{eqn:Smatrix2ch}
\end{equation}
where $\rho_{\rm{2NW}}=\sqrt{1-\left|\tau_{\rm{u} \sigma /\rm{d} \sigma}\right|^2}=\sqrt{1-\left|\tau'_{\rm{u} \sigma /\rm{d} \sigma}\right|^2}$;
$\tau_{\rm{u} \sigma /\rm{d} \sigma}\equiv t_2 t_{\rm{u} \sigma /\rm{d} \sigma}$
and $\tau'_{\rm{u} \sigma /\rm{d} \sigma}\equiv t_2 t'_{\rm{u} \sigma /\rm{d} \sigma}$
are the transmission amplitudes through a single (up ($\rm{u}$) or down ($\rm{d}$)) wire,
with the parameter~$t_2$ ranging from 0 to 1, $0\le t_2\le1$.

All these S-matrices fulfill the unitary condition \mbox{$S^{\dag}S=1$}.

\end{widetext}


\begin{thebibliography}{64}%
\makeatletter
\providecommand \@ifxundefined [1]{%
 \@ifx{#1\undefined}
}%
\providecommand \@ifnum [1]{%
 \ifnum #1\expandafter \@firstoftwo
 \else \expandafter \@secondoftwo
 \fi
}%
\providecommand \@ifx [1]{%
 \ifx #1\expandafter \@firstoftwo
 \else \expandafter \@secondoftwo
 \fi
}%
\providecommand \natexlab [1]{#1}%
\providecommand \enquote  [1]{``#1''}%
\providecommand \bibnamefont  [1]{#1}%
\providecommand \bibfnamefont [1]{#1}%
\providecommand \citenamefont [1]{#1}%
\providecommand \href@noop [0]{\@secondoftwo}%
\providecommand \href [0]{\begingroup \@sanitize@url \@href}%
\providecommand \@href[1]{\@@startlink{#1}\@@href}%
\providecommand \@@href[1]{\endgroup#1\@@endlink}%
\providecommand \@sanitize@url [0]{\catcode `\\12\catcode `\$12\catcode
  `\&12\catcode `\#12\catcode `\^12\catcode `\_12\catcode `\%12\relax}%
\providecommand \@@startlink[1]{}%
\providecommand \@@endlink[0]{}%
\providecommand \url  [0]{\begingroup\@sanitize@url \@url }%
\providecommand \@url [1]{\endgroup\@href {#1}{\urlprefix }}%
\providecommand \urlprefix  [0]{URL }%
\providecommand \Eprint [0]{\href }%
\providecommand \doibase [0]{http://dx.doi.org/}%
\providecommand \selectlanguage [0]{\@gobble}%
\providecommand \bibinfo  [0]{\@secondoftwo}%
\providecommand \bibfield  [0]{\@secondoftwo}%
\providecommand \translation [1]{[#1]}%
\providecommand \BibitemOpen [0]{}%
\providecommand \bibitemStop [0]{}%
\providecommand \bibitemNoStop [0]{.\EOS\space}%
\providecommand \EOS [0]{\spacefactor3000\relax}%
\providecommand \BibitemShut  [1]{\csname bibitem#1\endcsname}%
\let\auto@bib@innerbib\@empty
\bibitem [{\citenamefont {Recher}\ \emph {et~al.}(2001)\citenamefont {Recher},
  \citenamefont {Sukhorukov},\ and\ \citenamefont {Loss}}]{RecherPRB01}%
  \BibitemOpen
  \bibfield  {author} {\bibinfo {author} {\bibfnamefont {P.}~\bibnamefont
  {Recher}}, \bibinfo {author} {\bibfnamefont {E.~V.}\ \bibnamefont
  {Sukhorukov}}, \ and\ \bibinfo {author} {\bibfnamefont {D.}~\bibnamefont
  {Loss}},\ }\href@noop {} {\bibfield  {journal} {\bibinfo  {journal} {Phys.
  Rev. B}\ }\textbf {\bibinfo {volume} {63}},\ \bibinfo {pages} {165314}
  (\bibinfo {year} {2001})}\BibitemShut {NoStop}%
\bibitem [{\citenamefont {Eldridge}\ \emph {et~al.}(2010)\citenamefont
  {Eldridge}, \citenamefont {Pala}, \citenamefont {Governale},\ and\
  \citenamefont {K\"onig}}]{EldridgePRB10}%
  \BibitemOpen
  \bibfield  {author} {\bibinfo {author} {\bibfnamefont {J.}~\bibnamefont
  {Eldridge}}, \bibinfo {author} {\bibfnamefont {M.~G.}\ \bibnamefont {Pala}},
  \bibinfo {author} {\bibfnamefont {M.}~\bibnamefont {Governale}}, \ and\
  \bibinfo {author} {\bibfnamefont {J.}~\bibnamefont {K\"onig}},\ }\href@noop
  {} {\bibfield  {journal} {\bibinfo  {journal} {Phys. Rev. B}\ }\textbf
  {\bibinfo {volume} {82}},\ \bibinfo {pages} {184507} (\bibinfo {year}
  {2010})}\BibitemShut {NoStop}%
\bibitem [{\citenamefont {Hofstetter}\ \emph {et~al.}(2009)\citenamefont
  {Hofstetter}, \citenamefont {Csonka}, \citenamefont {Nyg{\aa}rd},\ and\
  \citenamefont {Sch\"{o}nenberger}}]{HofstetterNature2009}%
  \BibitemOpen
  \bibfield  {author} {\bibinfo {author} {\bibfnamefont {L.}~\bibnamefont
  {Hofstetter}}, \bibinfo {author} {\bibfnamefont {S.}~\bibnamefont {Csonka}},
  \bibinfo {author} {\bibfnamefont {J.}~\bibnamefont {Nyg{\aa}rd}}, \ and\
  \bibinfo {author} {\bibfnamefont {C.}~\bibnamefont {Sch\"{o}nenberger}},\
  }\href@noop {} {\bibfield  {journal} {\bibinfo  {journal} {Nature}\ }\textbf
  {\bibinfo {volume} {461}},\ \bibinfo {pages} {960} (\bibinfo {year}
  {2009})}\BibitemShut {NoStop}%
\bibitem [{\citenamefont {Herrmann}\ \emph {et~al.}(2010)\citenamefont
  {Herrmann}, \citenamefont {Portier}, \citenamefont {Roche}, \citenamefont
  {Yeyati}, \citenamefont {Kontos},\ and\ \citenamefont
  {Strunk}}]{HerrmannPRL10}%
  \BibitemOpen
  \bibfield  {author} {\bibinfo {author} {\bibfnamefont {L.~G.}\ \bibnamefont
  {Herrmann}}, \bibinfo {author} {\bibfnamefont {F.}~\bibnamefont {Portier}},
  \bibinfo {author} {\bibfnamefont {P.}~\bibnamefont {Roche}}, \bibinfo
  {author} {\bibfnamefont {A.~L.}\ \bibnamefont {Yeyati}}, \bibinfo {author}
  {\bibfnamefont {T.}~\bibnamefont {Kontos}}, \ and\ \bibinfo {author}
  {\bibfnamefont {C.}~\bibnamefont {Strunk}},\ }\href@noop {} {\bibfield
  {journal} {\bibinfo  {journal} {Phys. Rev. Lett.}\ }\textbf {\bibinfo
  {volume} {104}},\ \bibinfo {pages} {026801} (\bibinfo {year}
  {2010})}\BibitemShut {NoStop}%
\bibitem [{\citenamefont {Schindele}\ \emph {et~al.}(2012)\citenamefont
  {Schindele}, \citenamefont {Baumgartner},\ and\ \citenamefont
  {Sch\"onenberger}}]{SchindelePRL12}%
  \BibitemOpen
  \bibfield  {author} {\bibinfo {author} {\bibfnamefont {J.}~\bibnamefont
  {Schindele}}, \bibinfo {author} {\bibfnamefont {A.}~\bibnamefont
  {Baumgartner}}, \ and\ \bibinfo {author} {\bibfnamefont {C.}~\bibnamefont
  {Sch\"onenberger}},\ }\href@noop {} {\bibfield  {journal} {\bibinfo
  {journal} {Phys. Rev. Lett.}\ }\textbf {\bibinfo {volume} {109}},\ \bibinfo
  {pages} {157002} (\bibinfo {year} {2012})}\BibitemShut {NoStop}%
\bibitem [{\citenamefont {Das}\ \emph {et~al.}(2012)\citenamefont {Das},
  \citenamefont {Ronen}, \citenamefont {Heiblum}, \citenamefont {Mahalu},
  \citenamefont {Kretinin},\ and\ \citenamefont {Shtrikman}}]{DasNC2012}%
  \BibitemOpen
  \bibfield  {author} {\bibinfo {author} {\bibfnamefont {A.}~\bibnamefont
  {Das}}, \bibinfo {author} {\bibfnamefont {Y.}~\bibnamefont {Ronen}}, \bibinfo
  {author} {\bibfnamefont {M.}~\bibnamefont {Heiblum}}, \bibinfo {author}
  {\bibfnamefont {D.}~\bibnamefont {Mahalu}}, \bibinfo {author} {\bibfnamefont
  {A.~V.}\ \bibnamefont {Kretinin}}, \ and\ \bibinfo {author} {\bibfnamefont
  {H.}~\bibnamefont {Shtrikman}},\ }\href@noop {} {\bibfield  {journal}
  {\bibinfo  {journal} {Nat. Comms.}\ }\textbf {\bibinfo {volume} {3}},\
  \bibinfo {pages} {1165} (\bibinfo {year} {2012})}\BibitemShut {NoStop}%
\bibitem [{\citenamefont {Tan}\ \emph {et~al.}(2015)\citenamefont {Tan},
  \citenamefont {Cox}, \citenamefont {Nieminen}, \citenamefont
  {L\"ahteenm\"aki}, \citenamefont {Golubev}, \citenamefont {Lesovik},\ and\
  \citenamefont {Hakonen}}]{TanPRL2015}%
  \BibitemOpen
  \bibfield  {author} {\bibinfo {author} {\bibfnamefont {Z.~B.}\ \bibnamefont
  {Tan}}, \bibinfo {author} {\bibfnamefont {D.}~\bibnamefont {Cox}}, \bibinfo
  {author} {\bibfnamefont {T.}~\bibnamefont {Nieminen}}, \bibinfo {author}
  {\bibfnamefont {P.}~\bibnamefont {L\"ahteenm\"aki}}, \bibinfo {author}
  {\bibfnamefont {D.}~\bibnamefont {Golubev}}, \bibinfo {author} {\bibfnamefont
  {G.~B.}\ \bibnamefont {Lesovik}}, \ and\ \bibinfo {author} {\bibfnamefont
  {P.~J.}\ \bibnamefont {Hakonen}},\ }\href@noop {} {\bibfield  {journal}
  {\bibinfo  {journal} {Phys. Rev. Lett.}\ }\textbf {\bibinfo {volume} {114}},\
  \bibinfo {pages} {096602} (\bibinfo {year} {2015})}\BibitemShut {NoStop}%
\bibitem [{\citenamefont {Nielsen}\ and\ \citenamefont
  {Chuang}(2000)}]{NielsenQI2000}%
  \BibitemOpen
  \bibfield  {author} {\bibinfo {author} {\bibfnamefont {M.~A.}\ \bibnamefont
  {Nielsen}}\ and\ \bibinfo {author} {\bibfnamefont {I.~L.}\ \bibnamefont
  {Chuang}},\ }\href@noop {} {\emph {\bibinfo {title} {Quantum Computation and
  Quantum Information (Cambridge Series on Information and the Natural
  Sciences)}}}\ (\bibinfo  {publisher} {Cambridge University Press},\ \bibinfo
  {year} {2000})\BibitemShut {NoStop}%
\bibitem [{\citenamefont {Deacon}\ \emph {et~al.}(2015)\citenamefont {Deacon},
  \citenamefont {Oiwa}, \citenamefont {Sailer}, \citenamefont {Baba},
  \citenamefont {Kanai}, \citenamefont {Shibata}, \citenamefont {Hirakawa},\
  and\ \citenamefont {Tarucha}}]{DeaconNC15}%
  \BibitemOpen
  \bibfield  {author} {\bibinfo {author} {\bibfnamefont {R.~S.}\ \bibnamefont
  {Deacon}}, \bibinfo {author} {\bibfnamefont {A.}~\bibnamefont {Oiwa}},
  \bibinfo {author} {\bibfnamefont {J.}~\bibnamefont {Sailer}}, \bibinfo
  {author} {\bibfnamefont {S.}~\bibnamefont {Baba}}, \bibinfo {author}
  {\bibfnamefont {Y.}~\bibnamefont {Kanai}}, \bibinfo {author} {\bibfnamefont
  {K.}~\bibnamefont {Shibata}}, \bibinfo {author} {\bibfnamefont
  {K.}~\bibnamefont {Hirakawa}}, \ and\ \bibinfo {author} {\bibfnamefont
  {S.}~\bibnamefont {Tarucha}},\ }\href@noop {} {\bibfield  {journal} {\bibinfo
   {journal} {Nature Communications}\ }\textbf {\bibinfo {volume} {6}},\
  \bibinfo {pages} {7446} (\bibinfo {year} {2015})}\BibitemShut {NoStop}%
\bibitem [{\citenamefont {Wang}\ and\ \citenamefont {Hu}(2011)}]{WangPRL11}%
  \BibitemOpen
  \bibfield  {author} {\bibinfo {author} {\bibfnamefont {Z.}~\bibnamefont
  {Wang}}\ and\ \bibinfo {author} {\bibfnamefont {X.}~\bibnamefont {Hu}},\
  }\href@noop {} {\bibfield  {journal} {\bibinfo  {journal} {Phys. Rev. Lett.}\
  }\textbf {\bibinfo {volume} {106}},\ \bibinfo {pages} {037002} (\bibinfo
  {year} {2011})}\BibitemShut {NoStop}%
\bibitem [{\citenamefont {Aharonov}\ and\ \citenamefont
  {Bohm}(1959)}]{AharonovPR59}%
  \BibitemOpen
  \bibfield  {author} {\bibinfo {author} {\bibfnamefont {Y.}~\bibnamefont
  {Aharonov}}\ and\ \bibinfo {author} {\bibfnamefont {D.}~\bibnamefont
  {Bohm}},\ }\href@noop {} {\bibfield  {journal} {\bibinfo  {journal} {Phys.
  Rev.}\ }\textbf {\bibinfo {volume} {115}},\ \bibinfo {pages} {485} (\bibinfo
  {year} {1959})}\BibitemShut {NoStop}%
\bibitem [{\citenamefont {Sharvin}\ and\ \citenamefont
  {Sharvin}(1981)}]{SharvinJETP81}%
  \BibitemOpen
  \bibfield  {author} {\bibinfo {author} {\bibfnamefont {D.~Y.}\ \bibnamefont
  {Sharvin}}\ and\ \bibinfo {author} {\bibfnamefont {Y.~V.}\ \bibnamefont
  {Sharvin}},\ }\href@noop {} {\bibfield  {journal} {\bibinfo  {journal} {JETP
  Letters}\ }\textbf {\bibinfo {volume} {34}},\ \bibinfo {pages} {272}
  (\bibinfo {year} {1981})}\BibitemShut {NoStop}%
\bibitem [{\citenamefont {Webb}\ \emph {et~al.}(1985)\citenamefont {Webb},
  \citenamefont {Washburn}, \citenamefont {Umbach},\ and\ \citenamefont
  {Laibowitz}}]{WebbPRL85}%
  \BibitemOpen
  \bibfield  {author} {\bibinfo {author} {\bibfnamefont {R.~A.}\ \bibnamefont
  {Webb}}, \bibinfo {author} {\bibfnamefont {S.}~\bibnamefont {Washburn}},
  \bibinfo {author} {\bibfnamefont {C.~P.}\ \bibnamefont {Umbach}}, \ and\
  \bibinfo {author} {\bibfnamefont {R.~B.}\ \bibnamefont {Laibowitz}},\
  }\href@noop {} {\bibfield  {journal} {\bibinfo  {journal} {Phys. Rev. Lett.}\
  }\textbf {\bibinfo {volume} {54}},\ \bibinfo {pages} {2696} (\bibinfo {year}
  {1985})}\BibitemShut {NoStop}%
\bibitem [{\citenamefont {van Oudenaarden}\ \emph {et~al.}(1998)\citenamefont
  {van Oudenaarden}, \citenamefont {Devoret}, \citenamefont {Nazarov},\ and\
  \citenamefont {Mooij}}]{vanOudenaardenNature98}%
  \BibitemOpen
  \bibfield  {author} {\bibinfo {author} {\bibfnamefont {A.}~\bibnamefont {van
  Oudenaarden}}, \bibinfo {author} {\bibfnamefont {M.~H.}\ \bibnamefont
  {Devoret}}, \bibinfo {author} {\bibfnamefont {Y.~V.}\ \bibnamefont
  {Nazarov}}, \ and\ \bibinfo {author} {\bibfnamefont {J.~E.}\ \bibnamefont
  {Mooij}},\ }\href@noop {} {\bibfield  {journal} {\bibinfo  {journal}
  {Nature}\ }\textbf {\bibinfo {volume} {391}},\ \bibinfo {pages} {768}
  (\bibinfo {year} {1998})}\BibitemShut {NoStop}%
\bibitem [{\citenamefont {Aharonov}\ and\ \citenamefont
  {Casher}(1984)}]{AharonovPRL84}%
  \BibitemOpen
  \bibfield  {author} {\bibinfo {author} {\bibfnamefont {Y.}~\bibnamefont
  {Aharonov}}\ and\ \bibinfo {author} {\bibfnamefont {A.}~\bibnamefont
  {Casher}},\ }\href@noop {} {\bibfield  {journal} {\bibinfo  {journal} {Phys.
  Rev. Lett.}\ }\textbf {\bibinfo {volume} {53}},\ \bibinfo {pages} {319}
  (\bibinfo {year} {1984})}\BibitemShut {NoStop}%
\bibitem [{\citenamefont {Cimmino}\ \emph {et~al.}(1989)\citenamefont
  {Cimmino}, \citenamefont {Opat}, \citenamefont {Klein}, \citenamefont
  {Kaiser}, \citenamefont {Werner}, \citenamefont {Arif},\ and\ \citenamefont
  {Clothier}}]{CimminoPRL89}%
  \BibitemOpen
  \bibfield  {author} {\bibinfo {author} {\bibfnamefont {A.}~\bibnamefont
  {Cimmino}}, \bibinfo {author} {\bibfnamefont {G.~I.}\ \bibnamefont {Opat}},
  \bibinfo {author} {\bibfnamefont {A.~G.}\ \bibnamefont {Klein}}, \bibinfo
  {author} {\bibfnamefont {H.}~\bibnamefont {Kaiser}}, \bibinfo {author}
  {\bibfnamefont {S.~A.}\ \bibnamefont {Werner}}, \bibinfo {author}
  {\bibfnamefont {M.}~\bibnamefont {Arif}}, \ and\ \bibinfo {author}
  {\bibfnamefont {R.}~\bibnamefont {Clothier}},\ }\href@noop {} {\bibfield
  {journal} {\bibinfo  {journal} {Phys. Rev. Lett.}\ }\textbf {\bibinfo
  {volume} {63}},\ \bibinfo {pages} {380} (\bibinfo {year} {1989})}\BibitemShut
  {NoStop}%
\bibitem [{\citenamefont {Mathur}\ and\ \citenamefont
  {Stone}(1992)}]{Mathur.68.2964}%
  \BibitemOpen
  \bibfield  {author} {\bibinfo {author} {\bibfnamefont {H.}~\bibnamefont
  {Mathur}}\ and\ \bibinfo {author} {\bibfnamefont {A.~D.}\ \bibnamefont
  {Stone}},\ }\href@noop {} {\bibfield  {journal} {\bibinfo  {journal} {Phys.
  Rev. Lett.}\ }\textbf {\bibinfo {volume} {68}},\ \bibinfo {pages} {2964}
  (\bibinfo {year} {1992})}\BibitemShut {NoStop}%
\bibitem [{\citenamefont {K\"onig}\ \emph {et~al.}(2006)\citenamefont
  {K\"onig}, \citenamefont {Tschetschetkin}, \citenamefont {Hankiewicz},
  \citenamefont {Sinova}, \citenamefont {Hock}, \citenamefont {Daumer},
  \citenamefont {Sch\"afer}, \citenamefont {Becker}, \citenamefont {Buhmann},\
  and\ \citenamefont {Molenkamp}}]{KonigPRL06}%
  \BibitemOpen
  \bibfield  {author} {\bibinfo {author} {\bibfnamefont {M.}~\bibnamefont
  {K\"onig}}, \bibinfo {author} {\bibfnamefont {A.}~\bibnamefont
  {Tschetschetkin}}, \bibinfo {author} {\bibfnamefont {E.~M.}\ \bibnamefont
  {Hankiewicz}}, \bibinfo {author} {\bibfnamefont {J.}~\bibnamefont {Sinova}},
  \bibinfo {author} {\bibfnamefont {V.}~\bibnamefont {Hock}}, \bibinfo {author}
  {\bibfnamefont {V.}~\bibnamefont {Daumer}}, \bibinfo {author} {\bibfnamefont
  {M.}~\bibnamefont {Sch\"afer}}, \bibinfo {author} {\bibfnamefont {C.~R.}\
  \bibnamefont {Becker}}, \bibinfo {author} {\bibfnamefont {H.}~\bibnamefont
  {Buhmann}}, \ and\ \bibinfo {author} {\bibfnamefont {L.~W.}\ \bibnamefont
  {Molenkamp}},\ }\href@noop {} {\bibfield  {journal} {\bibinfo  {journal}
  {Phys. Rev. Lett.}\ }\textbf {\bibinfo {volume} {96}},\ \bibinfo {pages}
  {076804} (\bibinfo {year} {2006})}\BibitemShut {NoStop}%
\bibitem [{\citenamefont {Bergsten}\ \emph {et~al.}(2006)\citenamefont
  {Bergsten}, \citenamefont {Kobayashi}, \citenamefont {Sekine},\ and\
  \citenamefont {Nitta}}]{BergstenPRL06}%
  \BibitemOpen
  \bibfield  {author} {\bibinfo {author} {\bibfnamefont {T.}~\bibnamefont
  {Bergsten}}, \bibinfo {author} {\bibfnamefont {T.}~\bibnamefont {Kobayashi}},
  \bibinfo {author} {\bibfnamefont {Y.}~\bibnamefont {Sekine}}, \ and\ \bibinfo
  {author} {\bibfnamefont {J.}~\bibnamefont {Nitta}},\ }\href@noop {}
  {\bibfield  {journal} {\bibinfo  {journal} {Phys. Rev. Lett.}\ }\textbf
  {\bibinfo {volume} {97}},\ \bibinfo {pages} {196803} (\bibinfo {year}
  {2006})}\BibitemShut {NoStop}%
\bibitem [{\citenamefont {Datta}\ and\ \citenamefont {Das}(1990)}]{DattaAPL90}%
  \BibitemOpen
  \bibfield  {author} {\bibinfo {author} {\bibfnamefont {S.}~\bibnamefont
  {Datta}}\ and\ \bibinfo {author} {\bibfnamefont {B.}~\bibnamefont {Das}},\
  }\href@noop {} {\bibfield  {journal} {\bibinfo  {journal} {Applied Physics
  Letters}\ }\textbf {\bibinfo {volume} {56}},\ \bibinfo {pages} {665}
  (\bibinfo {year} {1990})}\BibitemShut {NoStop}%
\bibitem [{\citenamefont {Koo}\ \emph {et~al.}(2009)\citenamefont {Koo},
  \citenamefont {Kwon}, \citenamefont {Eom}, \citenamefont {Chang},
  \citenamefont {Han},\ and\ \citenamefont {Johnson}}]{KooScience2009}%
  \BibitemOpen
  \bibfield  {author} {\bibinfo {author} {\bibfnamefont {H.~C.}\ \bibnamefont
  {Koo}}, \bibinfo {author} {\bibfnamefont {J.~H.}\ \bibnamefont {Kwon}},
  \bibinfo {author} {\bibfnamefont {J.}~\bibnamefont {Eom}}, \bibinfo {author}
  {\bibfnamefont {J.}~\bibnamefont {Chang}}, \bibinfo {author} {\bibfnamefont
  {S.~H.}\ \bibnamefont {Han}}, \ and\ \bibinfo {author} {\bibfnamefont
  {M.}~\bibnamefont {Johnson}},\ }\href@noop {} {\bibfield  {journal} {\bibinfo
   {journal} {Science}\ }\textbf {\bibinfo {volume} {325}},\ \bibinfo {pages}
  {1515} (\bibinfo {year} {2009})}\BibitemShut {NoStop}%
\bibitem [{\citenamefont {Rashba}(1960)}]{RashbaSPSS60}%
  \BibitemOpen
  \bibfield  {author} {\bibinfo {author} {\bibfnamefont {E.~I.}\ \bibnamefont
  {Rashba}},\ }\href@noop {} {\bibfield  {journal} {\bibinfo  {journal} {Sov.
  Phys. Solid. State}\ }\textbf {\bibinfo {volume} {2}},\ \bibinfo {pages}
  {1224} (\bibinfo {year} {1960})}\BibitemShut {NoStop}%
\bibitem [{\citenamefont {Bychkov}\ and\ \citenamefont
  {Rashba}(1984)}]{BychkovJPC84}%
  \BibitemOpen
  \bibfield  {author} {\bibinfo {author} {\bibfnamefont {Y.~A.}\ \bibnamefont
  {Bychkov}}\ and\ \bibinfo {author} {\bibfnamefont {E.~I.}\ \bibnamefont
  {Rashba}},\ }\href@noop {} {\bibfield  {journal} {\bibinfo  {journal} {J.
  Phys. C: Solid State Phys.}\ }\textbf {\bibinfo {volume} {17}},\ \bibinfo
  {pages} {6039} (\bibinfo {year} {1984})}\BibitemShut {NoStop}%
\bibitem [{\citenamefont {Manchon}\ \emph {et~al.}(2015)\citenamefont
  {Manchon}, \citenamefont {Koo}, \citenamefont {Nitta}, \citenamefont
  {Frolov},\ and\ \citenamefont {Duine}}]{ManchonNatMat2015}%
  \BibitemOpen
  \bibfield  {author} {\bibinfo {author} {\bibfnamefont {A.}~\bibnamefont
  {Manchon}}, \bibinfo {author} {\bibfnamefont {H.~C.}\ \bibnamefont {Koo}},
  \bibinfo {author} {\bibfnamefont {J.}~\bibnamefont {Nitta}}, \bibinfo
  {author} {\bibfnamefont {S.~M.}\ \bibnamefont {Frolov}}, \ and\ \bibinfo
  {author} {\bibfnamefont {R.~A.}\ \bibnamefont {Duine}},\ }\href@noop {}
  {\bibfield  {journal} {\bibinfo  {journal} {Nature Materials}\ }\textbf
  {\bibinfo {volume} {14}},\ \bibinfo {pages} {871} (\bibinfo {year}
  {2015})}\BibitemShut {NoStop}%
\bibitem [{RSO()}]{RSOI}%
  \BibitemOpen
  \href@noop {} {}\bibinfo {note} {{The Rashba spin-orbit
  interaction~\cite{RashbaSPSS60,BychkovJPC84,ManchonNatMat2015} can be
  described by the Hamiltonian \mbox{$H_{\rm{R}} =
  \frac{\eta}{\hbar}\left(\vec{p}\times\vec{\sigma}\right)_{\rm{y}}$}, where
  $\eta$ is the Rashba parameter and the $\rm{y}$ axis is perpendicular to the
  2DEG plane. If we restrict the movement of electrons to the $\rm{x}$
  direction ($k_{\rm{z}}=0$), due to the spin-orbit interaction electrons with
  the same energy and spin polarizations $\pm \rm{z}$ will have different wave
  vectors, $k_\uparrow \neq k_\downarrow$. This implies different phases of
  spin-up and spin-down ($\sigma=~\uparrow,\downarrow$) electrons,
  $\phi_{\sigma}=k_\sigma L = \phi_0 + \sigma \phi_{\rm{R}}$, where $L$ is the
  length of the transport channel. Omitting the common phase factor $\phi_0$,
  it can be shown that the phase of a moving electron depends on its spin
  $\left| \sigma \right \rangle_{\rm{z}} \rightarrow \exp
  \left(\sigma\phi_{\rm{R}}\right) \left| \sigma \right
  \rangle_{\rm{z}}$.}}\BibitemShut {Stop}%
\bibitem [{\citenamefont {Bezuglyi}\ \emph {et~al.}(2002)\citenamefont
  {Bezuglyi}, \citenamefont {Rozhavsky}, \citenamefont {Vagner},\ and\
  \citenamefont {Wyder}}]{BezuglyiPRB02}%
  \BibitemOpen
  \bibfield  {author} {\bibinfo {author} {\bibfnamefont {E.~V.}\ \bibnamefont
  {Bezuglyi}}, \bibinfo {author} {\bibfnamefont {A.~S.}\ \bibnamefont
  {Rozhavsky}}, \bibinfo {author} {\bibfnamefont {I.~D.}\ \bibnamefont
  {Vagner}}, \ and\ \bibinfo {author} {\bibfnamefont {P.}~\bibnamefont
  {Wyder}},\ }\href@noop {} {\bibfield  {journal} {\bibinfo  {journal} {Phys.
  Rev. B}\ }\textbf {\bibinfo {volume} {66}},\ \bibinfo {pages} {052508}
  (\bibinfo {year} {2002})}\BibitemShut {NoStop}%
\bibitem [{\citenamefont {Krive}\ \emph {et~al.}(2004)\citenamefont {Krive},
  \citenamefont {Gorelik}, \citenamefont {Shekhter},\ and\ \citenamefont
  {Jonson}}]{KriveLTP04}%
  \BibitemOpen
  \bibfield  {author} {\bibinfo {author} {\bibfnamefont {I.~V.}\ \bibnamefont
  {Krive}}, \bibinfo {author} {\bibfnamefont {L.~Y.}\ \bibnamefont {Gorelik}},
  \bibinfo {author} {\bibfnamefont {R.~I.}\ \bibnamefont {Shekhter}}, \ and\
  \bibinfo {author} {\bibfnamefont {M.}~\bibnamefont {Jonson}},\ }\href@noop {}
  {\bibfield  {journal} {\bibinfo  {journal} {Low Temperature Physics}\
  }\textbf {\bibinfo {volume} {30}},\ \bibinfo {pages} {398} (\bibinfo {year}
  {2004})}\BibitemShut {NoStop}%
\bibitem [{\citenamefont {Krive}\ \emph {et~al.}(2005)\citenamefont {Krive},
  \citenamefont {Kadigrobov}, \citenamefont {Shekhter},\ and\ \citenamefont
  {Jonson}}]{KrivePRB05}%
  \BibitemOpen
  \bibfield  {author} {\bibinfo {author} {\bibfnamefont {I.~V.}\ \bibnamefont
  {Krive}}, \bibinfo {author} {\bibfnamefont {A.~M.}\ \bibnamefont
  {Kadigrobov}}, \bibinfo {author} {\bibfnamefont {R.~I.}\ \bibnamefont
  {Shekhter}}, \ and\ \bibinfo {author} {\bibfnamefont {M.}~\bibnamefont
  {Jonson}},\ }\href@noop {} {\bibfield  {journal} {\bibinfo  {journal} {Phys.
  Rev. B}\ }\textbf {\bibinfo {volume} {71}},\ \bibinfo {pages} {214516}
  (\bibinfo {year} {2005})}\BibitemShut {NoStop}%
\bibitem [{\citenamefont {Dell'Anna}\ \emph {et~al.}(2007)\citenamefont
  {Dell'Anna}, \citenamefont {Zazunov}, \citenamefont {Egger},\ and\
  \citenamefont {Martin}}]{DellAnnaPRB07}%
  \BibitemOpen
  \bibfield  {author} {\bibinfo {author} {\bibfnamefont {L.}~\bibnamefont
  {Dell'Anna}}, \bibinfo {author} {\bibfnamefont {A.}~\bibnamefont {Zazunov}},
  \bibinfo {author} {\bibfnamefont {R.}~\bibnamefont {Egger}}, \ and\ \bibinfo
  {author} {\bibfnamefont {T.}~\bibnamefont {Martin}},\ }\href@noop {}
  {\bibfield  {journal} {\bibinfo  {journal} {Phys. Rev. B}\ }\textbf {\bibinfo
  {volume} {75}},\ \bibinfo {pages} {085305} (\bibinfo {year}
  {2007})}\BibitemShut {NoStop}%
\bibitem [{\citenamefont {Buzdin}(2008)}]{BuzdinPRL08}%
  \BibitemOpen
  \bibfield  {author} {\bibinfo {author} {\bibfnamefont {A.}~\bibnamefont
  {Buzdin}},\ }\href@noop {} {\bibfield  {journal} {\bibinfo  {journal} {Phys.
  Rev. Lett.}\ }\textbf {\bibinfo {volume} {101}},\ \bibinfo {pages} {107005}
  (\bibinfo {year} {2008})}\BibitemShut {NoStop}%
\bibitem [{\citenamefont {Reynoso}\ \emph {et~al.}(2008)\citenamefont
  {Reynoso}, \citenamefont {Usaj}, \citenamefont {Balseiro}, \citenamefont
  {Feinberg},\ and\ \citenamefont {Avignon}}]{ReynosoPRL08}%
  \BibitemOpen
  \bibfield  {author} {\bibinfo {author} {\bibfnamefont {A.~A.}\ \bibnamefont
  {Reynoso}}, \bibinfo {author} {\bibfnamefont {G.}~\bibnamefont {Usaj}},
  \bibinfo {author} {\bibfnamefont {C.~A.}\ \bibnamefont {Balseiro}}, \bibinfo
  {author} {\bibfnamefont {D.}~\bibnamefont {Feinberg}}, \ and\ \bibinfo
  {author} {\bibfnamefont {M.}~\bibnamefont {Avignon}},\ }\href@noop {}
  {\bibfield  {journal} {\bibinfo  {journal} {Phys. Rev. Lett.}\ }\textbf
  {\bibinfo {volume} {101}},\ \bibinfo {pages} {107001} (\bibinfo {year}
  {2008})}\BibitemShut {NoStop}%
\bibitem [{\citenamefont {Zazunov}\ \emph {et~al.}(2009)\citenamefont
  {Zazunov}, \citenamefont {Egger}, \citenamefont {Jonckheere},\ and\
  \citenamefont {Martin}}]{ZazunovPRL09}%
  \BibitemOpen
  \bibfield  {author} {\bibinfo {author} {\bibfnamefont {A.}~\bibnamefont
  {Zazunov}}, \bibinfo {author} {\bibfnamefont {R.}~\bibnamefont {Egger}},
  \bibinfo {author} {\bibfnamefont {T.}~\bibnamefont {Jonckheere}}, \ and\
  \bibinfo {author} {\bibfnamefont {T.}~\bibnamefont {Martin}},\ }\href@noop {}
  {\bibfield  {journal} {\bibinfo  {journal} {Phys. Rev. Lett.}\ }\textbf
  {\bibinfo {volume} {103}},\ \bibinfo {pages} {147004} (\bibinfo {year}
  {2009})}\BibitemShut {NoStop}%
\bibitem [{\citenamefont {Brunetti}\ \emph {et~al.}(2013)\citenamefont
  {Brunetti}, \citenamefont {Zazunov}, \citenamefont {Kundu},\ and\
  \citenamefont {Egger}}]{BrunettiPRB13}%
  \BibitemOpen
  \bibfield  {author} {\bibinfo {author} {\bibfnamefont {A.}~\bibnamefont
  {Brunetti}}, \bibinfo {author} {\bibfnamefont {A.}~\bibnamefont {Zazunov}},
  \bibinfo {author} {\bibfnamefont {A.}~\bibnamefont {Kundu}}, \ and\ \bibinfo
  {author} {\bibfnamefont {R.}~\bibnamefont {Egger}},\ }\href@noop {}
  {\bibfield  {journal} {\bibinfo  {journal} {Phys. Rev. B}\ }\textbf {\bibinfo
  {volume} {88}},\ \bibinfo {pages} {144515} (\bibinfo {year}
  {2013})}\BibitemShut {NoStop}%
\bibitem [{\citenamefont {Yokoyama}\ \emph {et~al.}(2014)\citenamefont
  {Yokoyama}, \citenamefont {Eto},\ and\ \citenamefont
  {Nazarov}}]{YokoyamaPRB14}%
  \BibitemOpen
  \bibfield  {author} {\bibinfo {author} {\bibfnamefont {T.}~\bibnamefont
  {Yokoyama}}, \bibinfo {author} {\bibfnamefont {M.}~\bibnamefont {Eto}}, \
  and\ \bibinfo {author} {\bibfnamefont {Y.~V.}\ \bibnamefont {Nazarov}},\
  }\href@noop {} {\bibfield  {journal} {\bibinfo  {journal} {Phys. Rev. B}\
  }\textbf {\bibinfo {volume} {89}},\ \bibinfo {pages} {195407} (\bibinfo
  {year} {2014})}\BibitemShut {NoStop}%
\bibitem [{\citenamefont {Samokhvalov}\ \emph {et~al.}(2014)\citenamefont
  {Samokhvalov}, \citenamefont {Shekhter},\ and\ \citenamefont
  {Buzdin}}]{SamokhvalovSR14}%
  \BibitemOpen
  \bibfield  {author} {\bibinfo {author} {\bibfnamefont {A.~V.}\ \bibnamefont
  {Samokhvalov}}, \bibinfo {author} {\bibfnamefont {R.~I.}\ \bibnamefont
  {Shekhter}}, \ and\ \bibinfo {author} {\bibfnamefont {A.~I.}\ \bibnamefont
  {Buzdin}},\ }\href@noop {} {\bibfield  {journal} {\bibinfo  {journal} {Sci.
  Rep.}\ }\textbf {\bibinfo {volume} {4}},\ \bibinfo {pages} {5671} (\bibinfo
  {year} {2014})}\BibitemShut {NoStop}%
\bibitem [{\citenamefont {Jacobsen}\ and\ \citenamefont
  {Linder}(2015)}]{JacobsenPRB15}%
  \BibitemOpen
  \bibfield  {author} {\bibinfo {author} {\bibfnamefont {S.~H.}\ \bibnamefont
  {Jacobsen}}\ and\ \bibinfo {author} {\bibfnamefont {J.}~\bibnamefont
  {Linder}},\ }\href@noop {} {\bibfield  {journal} {\bibinfo  {journal} {Phys.
  Rev. B}\ }\textbf {\bibinfo {volume} {92}},\ \bibinfo {pages} {024501}
  (\bibinfo {year} {2015})}\BibitemShut {NoStop}%
\bibitem [{\citenamefont {Mironov}\ \emph {et~al.}(2015)\citenamefont
  {Mironov}, \citenamefont {Mel'nikov},\ and\ \citenamefont
  {Buzdin}}]{MironovPRL15}%
  \BibitemOpen
  \bibfield  {author} {\bibinfo {author} {\bibfnamefont {S.~V.}\ \bibnamefont
  {Mironov}}, \bibinfo {author} {\bibfnamefont {A.~S.}\ \bibnamefont
  {Mel'nikov}}, \ and\ \bibinfo {author} {\bibfnamefont {A.~I.}\ \bibnamefont
  {Buzdin}},\ }\href@noop {} {\bibfield  {journal} {\bibinfo  {journal} {Phys.
  Rev. Lett.}\ }\textbf {\bibinfo {volume} {114}},\ \bibinfo {pages} {227001}
  (\bibinfo {year} {2015})}\BibitemShut {NoStop}%
\bibitem [{\citenamefont {Campagnano}\ \emph {et~al.}(2015)\citenamefont
  {Campagnano}, \citenamefont {Lucignano}, \citenamefont {Giuliano},\ and\
  \citenamefont {Tagliacozzo}}]{CampagnanoJPCM15}%
  \BibitemOpen
  \bibfield  {author} {\bibinfo {author} {\bibfnamefont {G.}~\bibnamefont
  {Campagnano}}, \bibinfo {author} {\bibfnamefont {P.}~\bibnamefont
  {Lucignano}}, \bibinfo {author} {\bibfnamefont {D.}~\bibnamefont {Giuliano}},
  \ and\ \bibinfo {author} {\bibfnamefont {A.}~\bibnamefont {Tagliacozzo}},\
  }\href@noop {} {\bibfield  {journal} {\bibinfo  {journal} {Journal of
  Physics: Condensed Matter}\ }\textbf {\bibinfo {volume} {27}},\ \bibinfo
  {pages} {205301} (\bibinfo {year} {2015})}\BibitemShut {NoStop}%
\bibitem [{\citenamefont {Beenakker}(1991)}]{BeenakkerPRL91}%
  \BibitemOpen
  \bibfield  {author} {\bibinfo {author} {\bibfnamefont {C.~W.~J.}\
  \bibnamefont {Beenakker}},\ }\href@noop {} {\bibfield  {journal} {\bibinfo
  {journal} {Phys. Rev. Lett.}\ }\textbf {\bibinfo {volume} {67}},\ \bibinfo
  {pages} {3836} (\bibinfo {year} {1991})}\BibitemShut {NoStop}%
\bibitem [{\citenamefont {Bagwell}(1992)}]{BagwellPRB92}%
  \BibitemOpen
  \bibfield  {author} {\bibinfo {author} {\bibfnamefont {P.~F.}\ \bibnamefont
  {Bagwell}},\ }\href@noop {} {\bibfield  {journal} {\bibinfo  {journal} {Phys.
  Rev. B}\ }\textbf {\bibinfo {volume} {46}},\ \bibinfo {pages} {12573}
  (\bibinfo {year} {1992})}\BibitemShut {NoStop}%
\bibitem [{\citenamefont {Lee}\ \emph {et~al.}(2013)\citenamefont {Lee},
  \citenamefont {Jiang}, \citenamefont {Houzet}, \citenamefont {Aguado},
  \citenamefont {Lieber},\ and\ \citenamefont
  {Franceschi}}]{LeeNatureNano2013}%
  \BibitemOpen
  \bibfield  {author} {\bibinfo {author} {\bibfnamefont {E.~J.~H.}\
  \bibnamefont {Lee}}, \bibinfo {author} {\bibfnamefont {X.}~\bibnamefont
  {Jiang}}, \bibinfo {author} {\bibfnamefont {M.}~\bibnamefont {Houzet}},
  \bibinfo {author} {\bibfnamefont {R.}~\bibnamefont {Aguado}}, \bibinfo
  {author} {\bibfnamefont {C.~M.}\ \bibnamefont {Lieber}}, \ and\ \bibinfo
  {author} {\bibfnamefont {S.~D.}\ \bibnamefont {Franceschi}},\ }\href@noop {}
  {\bibfield  {journal} {\bibinfo  {journal} {Nature Nanotechnology}\ }\textbf
  {\bibinfo {volume} {9}},\ \bibinfo {pages} {79} (\bibinfo {year}
  {2013})}\BibitemShut {NoStop}%
\bibitem [{\citenamefont {van Woerkom}\ \emph {et~al.}(2017)\citenamefont {van
  Woerkom}, \citenamefont {Proutski}, \citenamefont {van Heck}, \citenamefont
  {Bouman}, \citenamefont {V{\"a}yrynen}, \citenamefont {Glazman},
  \citenamefont {Krogstrup}, \citenamefont {Nyg{\aa}rd}, \citenamefont
  {Kouwenhoven},\ and\ \citenamefont {Geresdi}}]{WoerkomArxiv2016}%
  \BibitemOpen
  \bibfield  {author} {\bibinfo {author} {\bibfnamefont {D.~J.}\ \bibnamefont
  {van Woerkom}}, \bibinfo {author} {\bibfnamefont {A.}~\bibnamefont
  {Proutski}}, \bibinfo {author} {\bibfnamefont {B.}~\bibnamefont {van Heck}},
  \bibinfo {author} {\bibfnamefont {D.}~\bibnamefont {Bouman}}, \bibinfo
  {author} {\bibfnamefont {J.~I.}\ \bibnamefont {V{\"a}yrynen}}, \bibinfo
  {author} {\bibfnamefont {L.~I.}\ \bibnamefont {Glazman}}, \bibinfo {author}
  {\bibfnamefont {P.}~\bibnamefont {Krogstrup}}, \bibinfo {author}
  {\bibfnamefont {J.}~\bibnamefont {Nyg{\aa}rd}}, \bibinfo {author}
  {\bibfnamefont {L.~P.}\ \bibnamefont {Kouwenhoven}}, \ and\ \bibinfo {author}
  {\bibfnamefont {A.}~\bibnamefont {Geresdi}},\ }\href@noop {} {\bibfield
  {journal} {\bibinfo  {journal} {Nat. Phys.}\ }\textbf {\bibinfo {volume}
  {13}},\ \bibinfo {pages} {876} (\bibinfo {year} {2017})}\BibitemShut
  {NoStop}%
\bibitem [{\citenamefont {Nazarov}\ and\ \citenamefont
  {Blanter}(2009)}]{NazarovQT}%
  \BibitemOpen
  \bibfield  {author} {\bibinfo {author} {\bibfnamefont {Y.~V.}\ \bibnamefont
  {Nazarov}}\ and\ \bibinfo {author} {\bibfnamefont {Y.~M.}\ \bibnamefont
  {Blanter}},\ }\href@noop {} {\emph {\bibinfo {title} {Quantum Transport:
  Introduction to Nanoscience}}}\ (\bibinfo  {publisher} {Cambridge University
  Press},\ \bibinfo {year} {2009})\BibitemShut {NoStop}%
\bibitem [{ABA()}]{ABAsy}%
  \BibitemOpen
  \href@noop {} {}\bibinfo {note} {{We do not consider asymmetry in the AB
  phases for the both arms since it does not affect the Josephson critical
  current~$I_{\rm{C}}$.}}\BibitemShut {Stop}%
\bibitem [{Uni()}]{UnitCond}%
  \BibitemOpen
  \href@noop {} {}\bibinfo {note} {{All these S-matrices fulfill the unitarity
  condition~$S^{\dag}S=1$.}}\BibitemShut {Stop}%
\bibitem [{Not()}]{NoteSymT}%
  \BibitemOpen
  \href@noop {} {}\bibinfo {note} {{This case is considered in
  Ref.~\cite{LiuPRB09}, where one can apply directly
  Eq.~(\ref{eqn:energyDef}).}}\BibitemShut {Stop}%
\bibitem [{\citenamefont {Washburn}\ and\ \citenamefont
  {Webb}(1986)}]{WashburnAP86}%
  \BibitemOpen
  \bibfield  {author} {\bibinfo {author} {\bibfnamefont {S.}~\bibnamefont
  {Washburn}}\ and\ \bibinfo {author} {\bibfnamefont {R.~A.}\ \bibnamefont
  {Webb}},\ }\href@noop {} {\bibfield  {journal} {\bibinfo  {journal} {Advances
  in Physics}\ }\textbf {\bibinfo {volume} {35}},\ \bibinfo {pages} {375}
  (\bibinfo {year} {1986})}\BibitemShut {NoStop}%
\bibitem [{\citenamefont {Mourachkine}(2004)}]{MourachkineJS2004}%
  \BibitemOpen
  \bibfield  {author} {\bibinfo {author} {\bibfnamefont {A.}~\bibnamefont
  {Mourachkine}},\ }\href@noop {} {\bibfield  {journal} {\bibinfo  {journal}
  {Journal of Superconductivity}\ }\textbf {\bibinfo {volume} {17}},\ \bibinfo
  {pages} {711} (\bibinfo {year} {2004})}\BibitemShut {NoStop}%
\bibitem [{\citenamefont {{Falci, G.}}\ \emph {et~al.}(2001)\citenamefont
  {{Falci, G.}}, \citenamefont {{Feinberg, D.}},\ and\ \citenamefont {{Hekking,
  F. W. J.}}}]{FalciEuroLet2001}%
  \BibitemOpen
  \bibfield  {author} {\bibinfo {author} {\bibnamefont {{Falci, G.}}}, \bibinfo
  {author} {\bibnamefont {{Feinberg, D.}}}, \ and\ \bibinfo {author}
  {\bibnamefont {{Hekking, F. W. J.}}},\ }\href@noop {} {\bibfield  {journal}
  {\bibinfo  {journal} {Europhys. Lett.}\ }\textbf {\bibinfo {volume} {54}},\
  \bibinfo {pages} {255} (\bibinfo {year} {2001})}\BibitemShut {NoStop}%
\bibitem [{\citenamefont {Oh}\ and\ \citenamefont {Kim}(2005)}]{OhPRB2005}%
  \BibitemOpen
  \bibfield  {author} {\bibinfo {author} {\bibfnamefont {S.}~\bibnamefont
  {Oh}}\ and\ \bibinfo {author} {\bibfnamefont {J.}~\bibnamefont {Kim}},\
  }\href@noop {} {\bibfield  {journal} {\bibinfo  {journal} {Phys. Rev. B}\
  }\textbf {\bibinfo {volume} {71}},\ \bibinfo {pages} {144523} (\bibinfo
  {year} {2005})}\BibitemShut {NoStop}%
\bibitem [{\citenamefont {Beckmann}\ \emph {et~al.}(2004)\citenamefont
  {Beckmann}, \citenamefont {Weber},\ and\ \citenamefont
  {v.~L\"ohneysen}}]{BeckmannPRL2004}%
  \BibitemOpen
  \bibfield  {author} {\bibinfo {author} {\bibfnamefont {D.}~\bibnamefont
  {Beckmann}}, \bibinfo {author} {\bibfnamefont {H.~B.}\ \bibnamefont {Weber}},
  \ and\ \bibinfo {author} {\bibfnamefont {H.}~\bibnamefont {v.~L\"ohneysen}},\
  }\href@noop {} {\bibfield  {journal} {\bibinfo  {journal} {Phys. Rev. Lett.}\
  }\textbf {\bibinfo {volume} {93}},\ \bibinfo {pages} {197003} (\bibinfo
  {year} {2004})}\BibitemShut {NoStop}%
\bibitem [{\citenamefont {Cadden-Zimansky}\ \emph {et~al.}(2009)\citenamefont
  {Cadden-Zimansky}, \citenamefont {Wei},\ and\ \citenamefont
  {Chandrasekhar}}]{CaddenZimanskyNatPhys2009}%
  \BibitemOpen
  \bibfield  {author} {\bibinfo {author} {\bibfnamefont {P.}~\bibnamefont
  {Cadden-Zimansky}}, \bibinfo {author} {\bibfnamefont {J.}~\bibnamefont
  {Wei}}, \ and\ \bibinfo {author} {\bibfnamefont {V.}~\bibnamefont
  {Chandrasekhar}},\ }\href@noop {} {\bibfield  {journal} {\bibinfo  {journal}
  {Nature Physics}\ }\textbf {\bibinfo {volume} {5}},\ \bibinfo {pages} {393}
  (\bibinfo {year} {2009})}\BibitemShut {NoStop}%
\bibitem [{\citenamefont {Brauer}\ \emph {et~al.}(2010)\citenamefont {Brauer},
  \citenamefont {H\"ubler}, \citenamefont {Smetanin}, \citenamefont
  {Beckmann},\ and\ \citenamefont {v.~L\"ohneysen}}]{BrauerPRB2010}%
  \BibitemOpen
  \bibfield  {author} {\bibinfo {author} {\bibfnamefont {J.}~\bibnamefont
  {Brauer}}, \bibinfo {author} {\bibfnamefont {F.}~\bibnamefont {H\"ubler}},
  \bibinfo {author} {\bibfnamefont {M.}~\bibnamefont {Smetanin}}, \bibinfo
  {author} {\bibfnamefont {D.}~\bibnamefont {Beckmann}}, \ and\ \bibinfo
  {author} {\bibfnamefont {H.}~\bibnamefont {v.~L\"ohneysen}},\ }\href@noop {}
  {\bibfield  {journal} {\bibinfo  {journal} {Phys. Rev. B}\ }\textbf {\bibinfo
  {volume} {81}},\ \bibinfo {pages} {024515} (\bibinfo {year}
  {2010})}\BibitemShut {NoStop}%
\bibitem [{Ass()}]{Assym}%
  \BibitemOpen
  \href@noop {} {}\bibinfo {note} {{The asymmetry of $t_2$ in the two nanowires
  is of no importance for the main conclusions. A nonzero Rashba phase in only
  one nanowire, e.g. $\phi_{\rm{Ru}}\neq0$ and $\phi_{\rm{Rd}}=0$, will suffice
  for the effects discussed in this Section to occur.}}\BibitemShut {Stop}%
\bibitem [{Par()}]{Paramt}%
  \BibitemOpen
  \href@noop {} {}\bibinfo {note} {{Using the analogy between scattering by a
  ring and a single nanowire with two scatterers, we define the parameter $t_2$
  for a single nanowire as $t_2=t^2/(2-t^2)$, where $t$ denotes the
  transmission amplitude of each separate scatterer.}}\BibitemShut {Stop}%
\bibitem [{Art()}]{ArticleQD}%
  \BibitemOpen
  \href@noop {} {}\bibinfo {note} {{D. Tomaszewski \textit{et al.} (in
  preparation).}}\BibitemShut {Stop}%
\bibitem [{\citenamefont {Fasth}\ \emph {et~al.}(2007)\citenamefont {Fasth},
  \citenamefont {Fuhrer}, \citenamefont {Samuelson}, \citenamefont {Golovach},\
  and\ \citenamefont {Loss}}]{FasthPRL2007}%
  \BibitemOpen
  \bibfield  {author} {\bibinfo {author} {\bibfnamefont {C.}~\bibnamefont
  {Fasth}}, \bibinfo {author} {\bibfnamefont {A.}~\bibnamefont {Fuhrer}},
  \bibinfo {author} {\bibfnamefont {L.}~\bibnamefont {Samuelson}}, \bibinfo
  {author} {\bibfnamefont {V.~N.}\ \bibnamefont {Golovach}}, \ and\ \bibinfo
  {author} {\bibfnamefont {D.}~\bibnamefont {Loss}},\ }\href@noop {} {\bibfield
   {journal} {\bibinfo  {journal} {Phys. Rev. Lett.}\ }\textbf {\bibinfo
  {volume} {98}},\ \bibinfo {pages} {266801} (\bibinfo {year}
  {2007})}\BibitemShut {NoStop}%
\bibitem [{\citenamefont {Nadj-Perge}\ \emph {et~al.}(2012)\citenamefont
  {Nadj-Perge}, \citenamefont {Pribiag}, \citenamefont {van~den Berg},
  \citenamefont {Zuo}, \citenamefont {Plissard}, \citenamefont {Bakkers},
  \citenamefont {Frolov},\ and\ \citenamefont
  {Kouwenhoven}}]{NadjPergePRL2012}%
  \BibitemOpen
  \bibfield  {author} {\bibinfo {author} {\bibfnamefont {S.}~\bibnamefont
  {Nadj-Perge}}, \bibinfo {author} {\bibfnamefont {V.~S.}\ \bibnamefont
  {Pribiag}}, \bibinfo {author} {\bibfnamefont {J.~W.~G.}\ \bibnamefont
  {van~den Berg}}, \bibinfo {author} {\bibfnamefont {K.}~\bibnamefont {Zuo}},
  \bibinfo {author} {\bibfnamefont {S.~R.}\ \bibnamefont {Plissard}}, \bibinfo
  {author} {\bibfnamefont {E.~P. A.~M.}\ \bibnamefont {Bakkers}}, \bibinfo
  {author} {\bibfnamefont {S.~M.}\ \bibnamefont {Frolov}}, \ and\ \bibinfo
  {author} {\bibfnamefont {L.~P.}\ \bibnamefont {Kouwenhoven}},\ }\href@noop {}
  {\bibfield  {journal} {\bibinfo  {journal} {Phys. Rev. Lett.}\ }\textbf
  {\bibinfo {volume} {108}},\ \bibinfo {pages} {166801} (\bibinfo {year}
  {2012})}\BibitemShut {NoStop}%
\bibitem [{\citenamefont {Mourik}\ \emph {et~al.}(2012)\citenamefont {Mourik},
  \citenamefont {Zuo}, \citenamefont {Frolov}, \citenamefont {Plissard},
  \citenamefont {Bakkers},\ and\ \citenamefont
  {Kouwenhoven}}]{MourikScience2012}%
  \BibitemOpen
  \bibfield  {author} {\bibinfo {author} {\bibfnamefont {V.}~\bibnamefont
  {Mourik}}, \bibinfo {author} {\bibfnamefont {K.}~\bibnamefont {Zuo}},
  \bibinfo {author} {\bibfnamefont {S.~M.}\ \bibnamefont {Frolov}}, \bibinfo
  {author} {\bibfnamefont {S.~R.}\ \bibnamefont {Plissard}}, \bibinfo {author}
  {\bibfnamefont {E.~P. A.~M.}\ \bibnamefont {Bakkers}}, \ and\ \bibinfo
  {author} {\bibfnamefont {L.~P.}\ \bibnamefont {Kouwenhoven}},\ }\href@noop {}
  {\bibfield  {journal} {\bibinfo  {journal} {Science}\ }\textbf {\bibinfo
  {volume} {336}},\ \bibinfo {pages} {1003} (\bibinfo {year}
  {2012})}\BibitemShut {NoStop}%
\bibitem [{\citenamefont {Baba}\ \emph {et~al.}(2017)\citenamefont {Baba},
  \citenamefont {Matsuo}, \citenamefont {Kamata}, \citenamefont {Deacon},
  \citenamefont {Oiwa}, \citenamefont {Li}, \citenamefont {Xu},\ and\
  \citenamefont {Tarucha}}]{BabaArxiv2017}%
  \BibitemOpen
  \bibfield  {author} {\bibinfo {author} {\bibfnamefont {S.}~\bibnamefont
  {Baba}}, \bibinfo {author} {\bibfnamefont {S.}~\bibnamefont {Matsuo}},
  \bibinfo {author} {\bibfnamefont {H.}~\bibnamefont {Kamata}}, \bibinfo
  {author} {\bibfnamefont {R.~S.}\ \bibnamefont {Deacon}}, \bibinfo {author}
  {\bibfnamefont {A.}~\bibnamefont {Oiwa}}, \bibinfo {author} {\bibfnamefont
  {K.}~\bibnamefont {Li}}, \bibinfo {author} {\bibfnamefont {H.~Q.}\
  \bibnamefont {Xu}}, \ and\ \bibinfo {author} {\bibfnamefont {S.}~\bibnamefont
  {Tarucha}},\ }\href@noop {} {}\bibinfo {howpublished} {arXiv:1703.03559}
  (\bibinfo {year} {2017})\BibitemShut {NoStop}%
\bibitem [{\citenamefont {Szombati}\ \emph {et~al.}(2016)\citenamefont
  {Szombati}, \citenamefont {Nadj-Perge}, \citenamefont {Car}, \citenamefont
  {Plissard}, \citenamefont {Bakkers},\ and\ \citenamefont
  {Kouwenhoven}}]{SzombatiNaturePhys2016}%
  \BibitemOpen
  \bibfield  {author} {\bibinfo {author} {\bibfnamefont {D.~B.}\ \bibnamefont
  {Szombati}}, \bibinfo {author} {\bibfnamefont {S.}~\bibnamefont
  {Nadj-Perge}}, \bibinfo {author} {\bibfnamefont {D.}~\bibnamefont {Car}},
  \bibinfo {author} {\bibfnamefont {S.~R.}\ \bibnamefont {Plissard}}, \bibinfo
  {author} {\bibfnamefont {E.~P. A.~M.}\ \bibnamefont {Bakkers}}, \ and\
  \bibinfo {author} {\bibfnamefont {L.~P.}\ \bibnamefont {Kouwenhoven}},\
  }\href@noop {} {\bibfield  {journal} {\bibinfo  {journal} {Nature Physics}\
  }\textbf {\bibinfo {volume} {12}},\ \bibinfo {pages} {568} (\bibinfo {year}
  {2016})}\BibitemShut {NoStop}%
\bibitem [{\citenamefont {Gazibegovic}\ \emph {et~al.}(2017)\citenamefont
  {Gazibegovic}, \citenamefont {Car}, \citenamefont {Zhang}, \citenamefont
  {Balk}, \citenamefont {Logan}, \citenamefont {de~Moor}, \citenamefont
  {Cassidy}, \citenamefont {Schmits}, \citenamefont {Xu}, \citenamefont {Wang},
  \citenamefont {Krogstrup}, \citenamefont {Op~het Veld}, \citenamefont {Zuo},
  \citenamefont {Vos}, \citenamefont {Shen}, \citenamefont {Bouman},
  \citenamefont {Shojaei}, \citenamefont {Pennachio}, \citenamefont {Lee},
  \citenamefont {van Veldhoven}, \citenamefont {Koelling}, \citenamefont
  {Verheijen}, \citenamefont {Kouwenhoven}, \citenamefont {Palmstr{\o}m},\ and\
  \citenamefont {Bakkers}}]{GazibegovicNature2017}%
  \BibitemOpen
  \bibfield  {author} {\bibinfo {author} {\bibfnamefont {S.}~\bibnamefont
  {Gazibegovic}}, \bibinfo {author} {\bibfnamefont {D.}~\bibnamefont {Car}},
  \bibinfo {author} {\bibfnamefont {H.}~\bibnamefont {Zhang}}, \bibinfo
  {author} {\bibfnamefont {S.~C.}\ \bibnamefont {Balk}}, \bibinfo {author}
  {\bibfnamefont {J.~A.}\ \bibnamefont {Logan}}, \bibinfo {author}
  {\bibfnamefont {M.~W.~A.}\ \bibnamefont {de~Moor}}, \bibinfo {author}
  {\bibfnamefont {M.~C.}\ \bibnamefont {Cassidy}}, \bibinfo {author}
  {\bibfnamefont {R.}~\bibnamefont {Schmits}}, \bibinfo {author} {\bibfnamefont
  {D.}~\bibnamefont {Xu}}, \bibinfo {author} {\bibfnamefont {G.}~\bibnamefont
  {Wang}}, \bibinfo {author} {\bibfnamefont {P.}~\bibnamefont {Krogstrup}},
  \bibinfo {author} {\bibfnamefont {R.~L.~M.}\ \bibnamefont {Op~het Veld}},
  \bibinfo {author} {\bibfnamefont {K.}~\bibnamefont {Zuo}}, \bibinfo {author}
  {\bibfnamefont {Y.}~\bibnamefont {Vos}}, \bibinfo {author} {\bibfnamefont
  {J.}~\bibnamefont {Shen}}, \bibinfo {author} {\bibfnamefont {D.}~\bibnamefont
  {Bouman}}, \bibinfo {author} {\bibfnamefont {B.}~\bibnamefont {Shojaei}},
  \bibinfo {author} {\bibfnamefont {D.}~\bibnamefont {Pennachio}}, \bibinfo
  {author} {\bibfnamefont {J.~S.}\ \bibnamefont {Lee}}, \bibinfo {author}
  {\bibfnamefont {P.~J.}\ \bibnamefont {van Veldhoven}}, \bibinfo {author}
  {\bibfnamefont {S.}~\bibnamefont {Koelling}}, \bibinfo {author}
  {\bibfnamefont {M.~A.}\ \bibnamefont {Verheijen}}, \bibinfo {author}
  {\bibfnamefont {L.~P.}\ \bibnamefont {Kouwenhoven}}, \bibinfo {author}
  {\bibfnamefont {C.~J.}\ \bibnamefont {Palmstr{\o}m}}, \ and\ \bibinfo
  {author} {\bibfnamefont {E.~P. A.~M.}\ \bibnamefont {Bakkers}},\ }\href@noop
  {} {\bibfield  {journal} {\bibinfo  {journal} {Nature}\ }\textbf {\bibinfo
  {volume} {548}},\ \bibinfo {pages} {434} (\bibinfo {year} {2017})},\ \bibinfo
  {note} {letter}\BibitemShut {NoStop}%
\bibitem [{\citenamefont {Ouisse}(2008)}]{Ouisse08}%
  \BibitemOpen
  \bibfield  {author} {\bibinfo {author} {\bibfnamefont {T.}~\bibnamefont
  {Ouisse}},\ }\href@noop {} {\emph {\bibinfo {title} {Electron Transport in
  Nanostructures and Mesoscopic Devices: An Introduction (ISTE)}}}\ (\bibinfo
  {publisher} {Wiley-ISTE},\ \bibinfo {year} {2008})\BibitemShut {NoStop}%
\bibitem [{\citenamefont {Liu}\ \emph {et~al.}(2009)\citenamefont {Liu},
  \citenamefont {Borunda}, \citenamefont {Liu},\ and\ \citenamefont
  {Sinova}}]{LiuPRB09}%
  \BibitemOpen
  \bibfield  {author} {\bibinfo {author} {\bibfnamefont {X.}~\bibnamefont
  {Liu}}, \bibinfo {author} {\bibfnamefont {M.~F.}\ \bibnamefont {Borunda}},
  \bibinfo {author} {\bibfnamefont {X.-J.}\ \bibnamefont {Liu}}, \ and\
  \bibinfo {author} {\bibfnamefont {J.}~\bibnamefont {Sinova}},\ }\href@noop {}
  {\bibfield  {journal} {\bibinfo  {journal} {Phys. Rev. B}\ }\textbf {\bibinfo
  {volume} {80}},\ \bibinfo {pages} {174524} (\bibinfo {year}
  {2009})}\BibitemShut {NoStop}%
\end{thebibliography}
\end{document}